\documentclass[12pt]{article}

\usepackage[dvips]{graphicx}
\usepackage{epsfig}
\usepackage{amsmath,amsfonts,amssymb,amsthm}
\usepackage{mathrsfs,mathtools}
\usepackage{verbatim}
\usepackage{psfrag}
\usepackage{bm}
\usepackage{bbm}
\usepackage[utf8]{inputenc}
\usepackage[square,comma,sort&compress,numbers]{natbib}
\usepackage[dvipsnames]{xcolor}
\usepackage{slashed}
\usepackage{upgreek}
\usepackage[normalem]{ulem}
\usepackage{enumitem}
\usepackage{float}
\usepackage[T1]{fontenc}
\usepackage{soul}
\usepackage{pgfplots}
\usepackage{extarrows}
\pgfplotsset{compat=1.17}
\usetikzlibrary{external,decorations.pathmorphing}
\usepackage{cases}
\usepackage{listings}

\lstdefinestyle{mystyle}{
    basicstyle=\ttfamily\footnotesize,
    breakatwhitespace=false,         
    breaklines=true,                 
    captionpos=b,                    
    keepspaces=true,                 
    numbers=left,                    
    numbersep=5pt,                  
    showspaces=false,                
    showstringspaces=false,
    showtabs=false,                  
    tabsize=2
}
\usepackage{lmodern}
\usepackage{tikz}

\usepackage[T1]{fontenc}

\tikzset{
  external/only named=true,
  thick/.style={line width=.5pt},
  approximation/.style={line width=1.2pt},
  numerics/.style={black, dotted, line width=.8pt},
  amplitude/.style={dashed},
  estimate/.style={dashed, line width=.8pt},
  normal plot/.style={line width=.8pt},
}

\tikzset{snake it/.style={decorate, decoration=snake}}

\usepackage[colorlinks=true,urlcolor=blue,anchorcolor=blue,citecolor=blue,filecolor=blue
,linkcolor=blue,menucolor=blue,linktocpage=true,pdfproducer=medialab,pdfa=true
]{hyperref}

\usepackage{epsf,epsfig}
\usepackage{graphics}
\usepackage{subcaption}

\newcommand\blfootnote[1]{%
  \begingroup
  \renewcommand\thefootnote{}\footnote{#1}%
  \addtocounter{footnote}{-1}%
  \endgroup
}

\def\d{\mathrm{d}}

\def\L{\mathcal{L}}

\def\O{\mathcal{O}}
\def\vec{\mathbf}
\def\i{\mathrm{i}}
\def\e{\mathrm{e}}
\def\P{\mathcal{P}}

\def\A{\mathcal{A}}

\def\M{\mathcal{M}}
\def\veck{\vec{k}}
\def\vecp{\vec{p}}

\def\vecx{\vec{x}}

\def\hata{\hat{a}}

\usepackage[errorstop]{feynmp}
\newcommand{\lsim}
{\;\raisebox{-.3em}{$\stackrel{\displaystyle <}{\sim}$}\;}
\newcommand{\gsim}
{\;\raisebox{-.3em}{$\stackrel{\displaystyle >}{\sim}$}\;}

\newcommand{\rom}[1]{\uppercase\expandafter{\romannumeral #1\relax}}

\begin{document}                                                                                                                                                                                                                                                                

\thispagestyle{empty}

\begin{flushright}
{
\small
KCL-PH-TH/2023-45
}
\end{flushright}

\vspace{-0.5cm}

\begin{center}
\Large\bf\boldmath
Logarithmically divergent friction on\\ ultrarelativistic bubble walls
\unboldmath
\end{center}

\vspace{-0.2cm}

\begin{center}
Wen-Yuan Ai$^{*1}$\blfootnote{$^*$wenyuan.ai@kcl.ac.uk} \\
\vskip0.4cm

{\it $^1$Theoretical Particle Physics and Cosmology, King’s College London,\\ Strand, London WC2R 2LS, United Kingdom} \\
\vskip1.cm
\end{center}

\begin{abstract}
We calculate the friction experienced by ultrarelativistic bubble walls resulting from the $1 \rightarrow 2$ light-to-heavy transition process, with finite-wall-width effects fully taken into account. In this process, the light particle is excited from the order-parameter scalar field, while the two heavy particles are excitations of a dark matter scalar field. Unlike earlier estimates suggesting a friction scaling as $\gamma_w^0$, where $\gamma_w$ represents the Lorentz factor of the wall velocity, our more precise numerical analysis reveals a logarithmic dependence of the friction on $\gamma_w$. We offer a numerical fit to capture this frictional pressure accurately. Our analysis verifies that the friction stemming from the $1 \rightarrow 2$ light-to-heavy transition is typically much smaller than the friction from the $1 \rightarrow 1$ transmission of the dark matter particles.
\end{abstract}

\newpage

\hrule
\tableofcontents
\vskip .85cm
\hrule


\section{Introduction}
\label{sec:Intro}

First-order phase transitions (FOPTs) remain to be a very intriguing possibility in the early Universe. Not only can such a first-order phase transition potentially play a significant role in generating the cosmic matter-antimatter asymmetry~\cite{Kuzmin:1985mm, Morrissey:2012db,Garbrecht:2018mrp}, but it also has the capacity to generate a stochastic gravitational wave background (SGWB)~\cite{Witten:1984rs,Kosowsky:1991ua,Kosowsky:1992vn,Kamionkowski:1993fg,Huber:2008hg,Hindmarsh:2013xza}.  This SGWB could be detectable through the forthcoming generation of gravitational wave (GW) detectors~\cite{Grojean:2006bp, Caprini:2018mtu, Mazumdar:2018dfl, Caprini:2019egz, LISACosmologyWorkingGroup:2022jok, Badger:2022nwo,Athron:2023xlk}. Very recently, several Pulsar Timing Array projects~\cite{NANOGrav:2023gor,EPTA:2023fyk,Reardon:2023gzh,Xu:2023wog} have reported the discovery of such an SGWB.

In the various phenomena linked to cosmological first-order phase transitions, the terminal bubble wall velocity is a key parameter. It impacts the magnitude of the matter-antimatter asymmetry in scenarios of FOPT-related baryogenesis~\cite{Cline:2020jre, Cline:2021dkf,Ellis:2022lft} and the shape and amplitude of the generated GW signals~\cite{Caprini:2015zlo,Gowling:2021gcy}. Estimates of the wall velocity are usually based on kinetic theory, where the coupled equations of motion for both the scalar field and plasma are solved, accounting for out-of-equilibrium effects~\cite{Liu:1992tn,Moore:1995ua,Moore:1995si,Konstandin:2014zta}. Alternatively, one adds to the equation of motion of the scalar field in thermal equilibrium an effective friction term~\cite{Ignatius:1993qn,Heckler:1994uu,Kurki-Suonio:1996gkq,Megevand:2009ut,Megevand:2009gh,Espinosa:2010hh}, whose coefficient can then be fixed by calculations in kinetic theory~\cite{Moore:2000wx,John:2000zq,Huber:2011aa,Huber:2013kj}. Typically, a full computation of the wall velocity is quite challenging and has only been performed for a  limited number of models~\cite{Liu:1992tn,Moore:1995si,Moore:1995ua,Dorsch:2018pat,Wang:2020zlf,Laurent:2022jrs,Jiang:2022btc}. 

Two distinct limiting cases significantly simplify the analysis. The first case arises when the plasma can be approximated to be in local thermal equilibrium~\cite{BarrosoMancha:2020fay,Balaji:2020yrx,Ai:2021kak,Ai:2023see} (see also Ref.~\cite{Wang:2022txy}). In this case, the plasma can still exert a backreaction force on the bubble wall provided the temperature is not homogeneous across the wall~\cite{Ignatius:1993qn,Konstandin:2010dm,Balaji:2020yrx,Ai:2021kak}. Notably, Refs.~\cite{Ai:2021kak,Ai:2023see} show that entropy conservation in local thermal equilibrium provides a new matching condition that can be used to determine the wall velocity efficiently. The second limiting case emerges when the wall is ultrarelativistic~\cite{Bodeker:2009qy,Bodeker:2017cim,Hoche:2020ysm,Azatov:2020ufh,Gouttenoire:2021kjv}. In this case, one can use the ballistic approximation~\cite{Dine:1992wr,Moore:1995si} and consider only particle transmission from the outside to the inside of the wall. While the second case is confined to extremely high Lorentz factors of the wall velocity, an investigation into it can aid in establishing whether the bubble wall runs away. Additionally, relativistic bubble walls offer a plethora of phenomenological implications~\cite{Azatov:2021ifm, Azatov:2021irb, Baldes:2021vyz, Huang:2022vkf, Azatov:2022tii, Baldes:2023fsp}. Thus, gaining a deeper understanding of the dynamics of bubble walls within this relativistic regime holds significant value. For other recent studies related to bubble growth, see Refs.~\cite{Friedlander:2020tnq,Cai:2020djd,Cline:2021iff,Bea:2021zsu,Bigazzi:2021ucw,Lewicki:2021pgr,Dorsch:2021nje,DeCurtis:2022hlx,Lewicki:2022nba,Ai:2022kqm,GarciaGarcia:2022yqb,LiLi:2023dlc,Krajewski:2023clt,Giombi:2023jqq}. 

In this paper, we analyze the second limiting case. It was first observed by Bodeker and Moore~\cite{Bodeker:2009qy} that the $1\rightarrow 1$ transmission of particles acquiring mass through the interaction with the order-parameter scalar contributes to a $\gamma_w$-independent backward pressure on the wall,
\begin{align}
    \mathcal{P}_{1\rightarrow 1}\simeq \sum_i g_i c_i \frac{\Delta m^2_i  T^2}{24}\,,
\end{align}
where $c_i=1(1/2)$ for bosons (fermions), $g_i$ is the number of internal degrees of freedom, $T$ is the nucleation temperature, and $\Delta m_i$ is the mass gain of the particle as it transitions from the exterior to the interior of the bubble. Later, the same authors realized that $1\rightarrow 2$ transition radiation, wherein a particle hitting the wall emits a soft gauge boson, contributes to a friction proportional to $\gamma_w$~\cite{Bodeker:2017cim}. This is given by
\begin{align}
    \mathcal{P}_{1\rightarrow 2;\rm  TR} \propto \gamma_w g^2 m_V T^3\,,
\end{align}
where $g$ is the gauge coupling, $m_V$ is the mass of the gauge boson.
Although Ref.~\cite{Hoche:2020ysm} argues that a resummation of multi-boson emissions enhances the power on $\gamma_w$ from one to two, Ref.~\cite{Gouttenoire:2021kjv} derives the same result as Bodeker and Moore's, with an additional factor logarithmic in $m_V$ found.\footnote{Ref.~\cite{BarrosoMancha:2020fay} claims that there is a $\gamma_w^2$-scaled friction in local thermal equilibrium. However, this may be due to an improper approximation in the plasma temperature and velocity used there, as pointed out in Ref.~\cite{Ai:2021kak}.}

Certainly, other types of $1\rightarrow 2$ processes would also result in a frictional pressure on the bubble wall.  In particular, the so-called light-to-heavy transition, $h\rightarrow 2\phi$, where $h$ is the scalar field responsible for the FOPT (after the spontaneous symmetry breaking) and $\phi$ is a heavy scalar dark matter field, has been analyzed in Refs.~\cite{Azatov:2020ufh,Azatov:2021ifm}. The Lagrangian term responsible for this process is
\begin{align}
\label{eq:Lagrangian}
    -\L\supset \frac{\lambda}{2} v(x) h \phi^2\,,
\end{align}
which is obtained from an interaction term $\lambda H^2 \phi^2/4$ via the spontaneous symmetry breaking $H(x)=\langle H(x)\rangle + h(x)\equiv v(x)+h(x)$.\footnote{Here, $v(x)$, in the general context, is called a background field or condensate. Except for particle production from fast-moving bubble walls, condensate-dependent interactions can also induce particle production for oscillating homogeneous configurations~\cite{Paz:1990sd,Boyanovsky:1994me,Yokoyama:2004pf,Mukaida:2012qn,Mukaida:2013xxa,Ai:2021gtg,Wang:2022mvv,Ai:2023ahr}, or in collisions of bubble walls~\cite{Watkins:1991zt,Konstandin:2011ds,Falkowski:2012fb,Mansour:2023fwj}.} The analysis in Refs.~\cite{Azatov:2020ufh,Azatov:2021ifm} suggests that the friction induced by this process is independent of $\gamma_w$. However, that analysis is based on an estimate with a few approximations made in their calculation. In this paper, we shall recompute this friction with more meticulous treatment. Of course, without those approximations made in Ref.~\cite{Azatov:2020ufh}, we cannot obtain an analytic expression. However, we will see that the exact numerical results of the pressure can be perfectly fitted by a logarithmic function in $\gamma_w$. (The dependence on $\gamma_w$ will be rather obvious; the question is whether this dependence goes away for $\gamma_w\rightarrow \infty$, which can happen via, e.g., $\exp(-c_1\times\gamma^{-c_2}_w)$ with $c_2>0$.)

The paper is organized as follows. In the next section, we derive the master integral for the backward friction in the ultrarelativistic regime, using the light-to-heavy transition as an example. We introduce the Fourier transform of the bubble profile in expressing the differential transition probability so that finite-wall-width effects are fully captured. In Sec.~\ref{sec:simplication}, we simplify the master integral. The simplified integral is then numerically computed in Sec.~\ref{sec:numerical-results} and a numerical fit is given for the results. We also comment on potential sources that might have caused the discrepancy between our numerical results and the analytic one in Ref.~\cite{Azatov:2020ufh}. We conclude in Sec.~\ref{sec:conclusion}.

\section{Friction from the light-to-heavy transition}
\label{sec:review}

\begin{figure}[ht]
    \centering
    \includegraphics[scale=0.45]{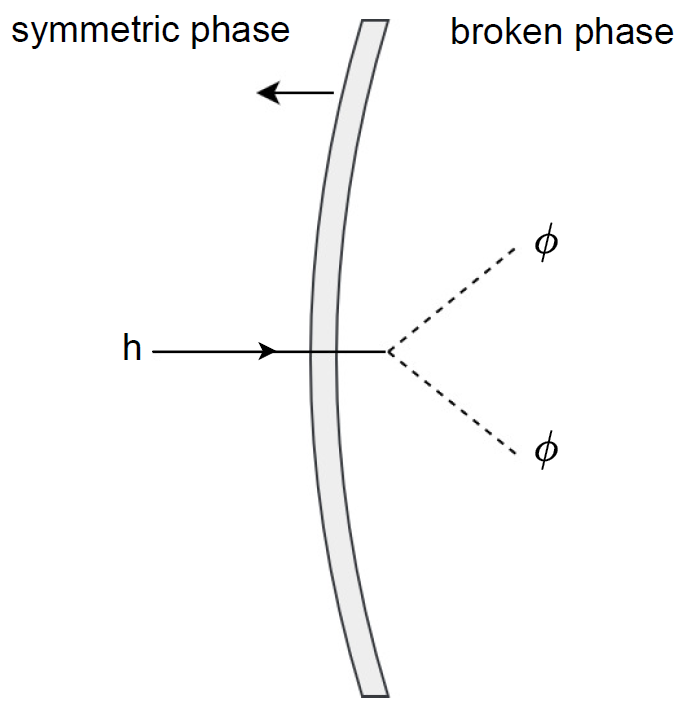}
    \caption{The light-to-heavy transition process. The incoming particle does not necessarily have a vanishing transverse momentum.}
    \label{fig:light-to-heavy}
\end{figure}

We assume a planar wall expanding in the negative $z$-direction and work in the rest frame of the wall. A conventional profile for the bubble wall is
\begin{align}
\label{eq:wall-profile}
    v(z)=\frac{v_b}{2}\left[{\rm tanh}(z/L_w)+1\right]\,,
\end{align}
where $v_b>0$ is the symmetry broken value, $L_w$ characterizes the wall width in the wall frame. Here we assume, without loss of generality, that the vacuum expectation value of $H$ vanishes in the symmetric phase. We consider the transition $h\rightarrow 2\phi$ described by Eq.~\eqref{eq:Lagrangian} and assume the following hierarchy for the mass scales
\begin{align}
\label{eq:hierachy}
    \gamma_w T\gg  m_\phi \gg  (T\sim v_b\sim m_h)\,,
\end{align}
where $m_\phi$ is the zero-temperature mass of $\phi$ before the spontaneous symmetry breaking. The light-to-heavy transition is schematically described in Fig.~\ref{fig:light-to-heavy}. This process is forbidden in the absence of the wall and becomes possible in the wall background since then the total $z$-momentum of particles is not required to be conserved.

Given our assumption that $m_\phi\gg (T \sim v_b\sim m_h)$, we can safely neglect the field-dependent and thermal masses for the $\phi$ particles. For the light-to-heavy process to occur, the $h$ particle must possess a very large $z$-momentum. Consequently, for these high-$z$-momentum modes, we can similarly neglect the field-dependent and thermal masses for $h$, treating them as effectively massless. We express the four-momenta as:
\begin{subequations}
\label{eq:kinematics}
\begin{align}
    &h:\quad\ \  p=(E^{(h)}_\vecp,\vecp_\perp,p^z)\,,\\
    &\phi_1:\quad k_1=(E^{(\phi)}_{\veck_1},\veck_{1,\perp},k_1^z)\,,\\
    &\phi_2:\quad k_2=(E^{(\phi)}_{\veck_2},\veck_{2,\perp},k_2^z)\,,
\end{align}
\end{subequations}
where $\vecp_\perp= (\vecp^x,\vecp^y)$ and similarly for $\veck_{1,\perp}$, $\veck_{2,\perp}$, and $E_\vecp^{(h)}=|\vecp|$, $E_\veck^{(\phi)}=\sqrt{m_\phi^2+\veck^2}$. 

The backward pressure on the wall due to the light-to-heavy transition is given by~\cite{Bodeker:2017cim}
\begin{align}
\label{eq:pressure}
    \mathcal{P}_{h\rightarrow 2\phi}=\int\frac{\d^3 \vecp}{(2\pi)^3}\frac{p^z}{E^{(h)}_\vecp} \d\mathbb{P}_{h\rightarrow 2\phi}(\vecp)\times f_h(\vecp,T)\times \Delta p^z\,,
\end{align}
where $\d\mathbb{P}_{h\rightarrow 2\phi}(\vecp)$ is the differential transition probability, $\Delta p^z=p^z-k_1^z-k_2^z$, and
\begin{align}
    f_h(\vecp,T)=\frac{1}{\e^{\frac{\gamma_w\left(E^{(h)}_\vecp-v_w p^z\right)}{T}}-1}\,,
\end{align}
with $\gamma_w \equiv 1/\sqrt{1-v_w^2}$.

\paragraph{Differential transition probability}

Let us first derive the expression of  $\d\mathbb{P}_{h\rightarrow 2\phi}(\vecp)$. The free $\phi$ field has the following decomposition
\begin{align}
\label{eq:decomp-phi}
    \phi_{\rm (free)}(x)= \int\frac{\d^3 \veck}{(2\pi)^3 2 E^{(\phi)}_{\veck}}  \left( \hata_\veck\e^{-\i(E^{(\phi)}_\veck t-\veck\cdot\vecx) }+ \hata_\veck^\dagger \e^{\i(E^{(\phi)}_\veck t-\veck\cdot\vecx )}\right)\,,
\end{align}
where $\hata_\veck^\dagger$, $\hata_\veck$ are the creation and annihilation operators satisfying
\begin{align}
    [\hata_\veck,\hata_{\veck'}]=0\,,\quad [\hata_\veck^\dagger,\hata_{\veck'}^\dagger]=0\,,\quad [\hata_\veck, \hata_{\veck'}^\dagger]=2E^{(\phi)}_\veck(2\pi)^3\delta^{(3)}(\veck-\veck')\,.
\end{align}
A one-particle state of $\phi$ is thus defined as
\begin{align}
    |\veck \rangle=\hata_\veck^\dagger |0\rangle\,,
\end{align}
and we have 
\begin{align}
    \langle \veck|\veck'\rangle =2E^{(\phi)}_\veck (2\pi)^3\delta^{(3)}(\veck-\veck')=2k^z (2\pi)^3 \delta(E_\veck^{(\phi)}-E_{\vecp'}^{\phi})\delta^{(2)}(\veck_\perp -\veck'_\perp)\,.
\end{align}

For the $h$ field, due to its mass being $z$-dependent, more careful treatment is usually required~\cite{Bodeker:2017cim}. However, as mentioned above, the process under study requires a very large $z$-momentum for the $h$-particle. As a consequence, one can still use the normal free field decomposition~\eqref{eq:decomp-phi} also for the $h$ (with the mass neglected). We use the momentum labels to distinguish single-particle states of $h$ and $\phi$, as indicated in Eq.~\eqref{eq:kinematics}, and write the one-particle state of $h$ simply as $|\vecp\rangle$. As usual, to study the decay probability of a $h$-particle, one needs to introduce its wave-packet state that can be properly normalized to one,
\begin{align}
\label{eq:wave-packet}
    |\psi_h(\vecp)\rangle =\int \frac{\d^3 \vecp'}{(2\pi)^3 2 E_{\vecp'}^{(h)}} \psi_h(\vecp') |\vecp'\rangle\,, \qquad {\rm with}\quad \int\frac{\d^3 \vecp}{(2\pi)^3 2E^{(h)}_\vecp}|\psi_h(\vecp)|^2=1\,.
\end{align}

Now one needs to study the following transition matrix element
\begin{align}
    \A_{h\rightarrow 2\phi}= \langle \veck_1\veck_2|(-\i) \int\d t\, H_{\rm int}|\psi_h(\vecp)\rangle =-\frac{\i \lambda}{2}\,\langle \veck_1\veck_2|\int\d^4 x\, v(z) h(x)\phi^2(x)|\psi_h(\vecp)\rangle\,.
\end{align}
The differential probability is then given by
\begin{align}
    \d\mathbb{P}_{h\rightarrow 2 \phi}(\vecp)=\prod_{i=1,2}\int\frac{\d^3\veck_i}{(2\pi)^3 2 E^{(\phi)}_{\veck_i}}\frac{1}{2}\times |\A_{h\rightarrow 2\phi}|^2\,,
\end{align}
where the particular factor of $1/2$  is due to the symmetry when exchanging two $\phi$ particles in the final state.
Substituting the decomposition~\eqref{eq:decomp-phi} for $\phi$ and $h$ into the above equation, using the commutation relations for the creation and annihilation operators and using Eq.~\eqref{eq:wave-packet}, one obtains
\begin{align}
\label{eq:propb-p}
    \d\mathbb{P}_{h\rightarrow 2\phi}(\vecp)=\frac{1}{4p^z}\prod_{i=1,2}\int\frac{\d^3\veck_i}{(2\pi)^3 2 E^{(\phi)}_{\veck_i}}  (2\pi)^3 \delta(E^{(h)}_{\vecp}-E^{(\phi)}_{\veck_1}-E^{(\phi)}_{\veck_2})\delta^{(2)}(\vecp_\perp-\veck_{1,\perp}-\veck_{2,\perp})|\M_{h\rightarrow 2\phi}|^2\,,
\end{align}
where
\begin{align}
     \i\M_{h\rightarrow 2\phi} =(-\i\lambda) \, \int\d z\, v(z)   \e^{\i  \Delta p^z z} \,.
\end{align}
Note that we used the same definition of $\M$ as in Ref.~\cite{Bodeker:2017cim} which differs from that used in Ref.~\cite{Hoche:2020ysm} by a factor of $p^z/E_\vecp^{(h)}$. The formula~\eqref{eq:propb-p} agrees with Refs.~\cite{Bodeker:2017cim,Hoche:2020ysm} except for the symmetry factor for the produced $\phi$ particles. When comparing the above formula with the one given in Ref.~\cite{Hoche:2020ysm}, one needs to keep in mind the difference in the definitions of $\M$.

Taking the Fourier transform of $v(z)$, 
\begin{align}
    v(z)=\int\frac{\d q^z}{2\pi}\e^{\i q^z z}\Tilde{v}(q^z)\,, 
\end{align}
in $\M_{h\rightarrow 2\phi}$, one obtains
\begin{align}
\label{eq:M-vtilde}
    \M_{h\rightarrow 2\phi}=-\lambda\, \Tilde{v}(k_{1}^z+k_2^z-p^z)\,.
\end{align}
This form of the invariant transition amplitude has a clear physical meaning; it describes the interaction between the three particles and a Fourier mode of the bubble wall with its four-momentum being $-\Delta p^z$. The total four-momentum of the particles {\it and} the wall is conserved.
For the profile~\eqref{eq:wall-profile}, we have
\begin{align}
\label{eq:vtilde}
    \tilde{v}(q^z)=\frac{\i v_b L_w \sqrt{\pi}}{2\
    \sqrt{2}} {\rm csch}\left(\frac{\pi L_w q^z}{2}\right)+\frac{v_b\sqrt{2\pi}}{2}\delta(q^z)\,.
\end{align}
The second term describes processes where there is no energy-momentum transfer between the wall and particles. However such processes cannot satisfy the on-shell conditions and hence the second term in $\tilde{v}(q^z)$ can be simply discarded.

Substituting Eqs.~\eqref{eq:M-vtilde} and~\eqref{eq:vtilde} into Eq.~\eqref{eq:propb-p}, one has 
\begin{align}
\label{eq:dP}
    \d\mathbb{P}_{h\rightarrow  2\phi}(\vec{p})=\frac{\lambda^2 v_b^2 L_{w}^2 \pi}{32 p^z}  &\prod_{i=1,2}\int\frac{\d^3\veck_i}{(2\pi)^3 2 E^{(\phi)}_{\veck_i}} \times (2\pi)^3 \delta(E^{(h)}_{\vecp}-E^{(\phi)}_{\veck_1}- E^{(\phi)}_{\veck_2})\delta^{(2)}(\vecp_\perp-\veck_{1,\perp}-\veck_{2,\perp})\notag\\
    &\times {\rm csch}^2\left(\frac{\pi L_w \Delta p^z }{2}\right) \,.
\end{align}

One can integrate over $k_1$ in the above equation using
\begin{align}
    \delta (E_{\veck_1}^{(\phi)}+E^{(\phi)}_{\veck_2}-E_\vecp^{(h)})=\frac{E_{\veck_1}^{(\phi)}}{k_1^z}\delta \left(k_1^z-\sqrt{H(\vecp,\veck_2)}\right)\,,
\end{align}
where 
\begin{align}
    H(\vecp,\veck_2)\equiv (p^z)^2+(k_2^z)^2+2\vecp_\perp\cdot \veck_{2,\perp}-2|\vecp|E_{\veck_2}^{(\phi)}\,.
\end{align}
After relabeling $\veck_2$ as $\veck$, we obtain
\begin{align}
\label{eq:dP-integrated-over-k1}
    \d \mathbb{P}_{h\rightarrow 2\phi}(\vecp)=\frac{\lambda^2 v_b^2 L_w^2 \pi }{64 p^z}\int\frac{\d^3 \veck}{(2\pi)^3 2E_\veck^{(\phi)}} \frac{1}{\sqrt{H(\vecp,\veck)}}\times {\rm csch}^2 \left(\frac{\pi L_w  \overline{\Delta p^z}}{2}\right)\,,
\end{align}
where
\begin{align}
    \overline{\Delta p^z} = p^z-k^z-\sqrt{H(\vecp,\veck)}\,.
\end{align}
In the above equations, there is the constraint $H(\vecp,\veck)\geq 0$ which is enforced by kinematics. To the best of our knowledge, the Fourier transform of the bubble profile has never been used in expressing the differential probability in the existing literature. Substituting Eq.~\eqref{eq:dP-integrated-over-k1} into
Eq.~\eqref{eq:pressure}, we obtain the following master integral,
\begin{align}
\label{eq:master-integral}
      \mathcal{P}_{h\rightarrow 2\phi}=\frac{\lambda^2 v_b^2 L_w^2 \pi }{64}\int\frac{\d^3 \vecp}{(2\pi)^3 E^{(h)}_\vecp} \int\frac{\d^3 \veck}{(2\pi)^3 2E_\veck^{(\phi)}} f_h(\vecp,T)\times \frac{\overline{\Delta p^z}}{\sqrt{H(\vecp,\veck)}}\times {\rm csch}^2 \left(\frac{\pi L_w  \overline{\Delta p^z}}{2}\right)\,.
\end{align}

\section{Simplify the integral}
\label{sec:simplication}

In this section, we simplify the integral~\eqref{eq:master-integral} to render it suitable for numerical computations. We do this through the following two steps:
\begin{itemize}
    \item[(i)] Do the integral over $p^z$ using the method of steepest descent, making use of the fact that the distribution function $f_h(\vecp,T)$ is exponentially suppressed away from the stationary point $p^z=\gamma_w|\vecp_\perp|$ (see below).
    \item[(ii)] Set a cutoff for $L_w\overline{\Delta p^z}$, making use of the fact the hyperbolic cosecant function is exponentially suppressed for large $L_w\overline{\Delta p^z}$. 
\end{itemize}

First, one can approximate the distribution function by the Boltzmann distribution, 
\begin{align}
    f_h(\vecp,T)\approx \e^{-[\gamma_w(E_\vecp^{(h)}-v_w p^z)]/T}\,.
\end{align}
If $p^z\gg |\vecp_\perp|$, we have
\begin{align}
    E_\vecp^{(h)}\approx p^z+\frac{1}{2}\frac{\vecp_\perp^2}{p^z}\,.
\end{align}
Substituting the above and $v_w\approx 1-1/(2\gamma_w^2)$ (since $\gamma_w \gg 1$) into $f_h(\vecp,T)$, one obtains 
\begin{align}
\label{eq:fh-bound}
    f_h(\vecp,T)\approx \e^{-\frac{1}{2T} \left(\frac{\gamma_w \vecp_\perp^2}{p^z}+\frac{p^z}{\gamma_w}\right)} \leq \e^{-\frac{|\vecp_\perp|}{T}} \,,
\end{align}
where the equality is saturated at $p^z=\gamma_w|\vecp_\perp|$. If, on the other hand, $p^z\lsim |\vecp_\perp|$, one can show that 
\begin{align}
    f_h(\vecp,T)\approx \e^{-\frac{\#\gamma_w |\vecp_\perp|}{T}}\,,
\end{align}
where $\#$ is a number roughly from $\sqrt{2}-1$ to $1$. The above result is exponentially suppressed compared with the bound given in Eq.~\eqref{eq:fh-bound}, so that the integral is dominated for $p^z\gg |\vecp_\perp|$.

Using Eq.~\eqref{eq:fh-bound}, one can therefore evaluate the integral over $p^z$ in Eq.~\eqref{eq:pressure} by the method of steepest descent (we only expand the exponent of $f_h$ to the second-order derivative while all others in the integrand are evaluated at the stationary point $p^z=\gamma_w|\vecp_\perp|$). We obtain
\begin{align}
    &\mathcal{P}_{h\rightarrow 2\phi}= \frac{1}{4\pi^2}\int\d |\vecp_\perp|\, |\vecp_\perp|\int \d p^z\, \frac{p^z}{E_\vecp^{(h)}}\,\d\mathbb{P}_{h\rightarrow 2\phi}(|\vecp_\perp|,p^z)\times f_h(\vecp,T)\times \overline{\Delta p^z}(|\vecp_\perp|,p^z,\veck)\notag\\
    &\ \approx  \frac{1}{4\pi^2}\int\d |\vecp_\perp|\, |\vecp_\perp| \times\left(\gamma_w\sqrt{2\pi T|\vecp_\perp|}\, \d\mathbb{P}_{h\rightarrow 2\phi}(|\vecp_\perp|,\gamma_w|\vecp_\perp|)\,\overline{\Delta p^z}(|\vecp_\perp|,\gamma_w|\vecp_\perp|,\veck)\,\e^{-\frac{|\vecp_\perp|}{T}}\right)\,,
\end{align}
where in the second step we have done the Gaussian integral and used $p^z\approx E_\vecp^{(h)}$ at the stationary point since $\gamma_w\gg 1$.

Now let us discuss $\d\mathbb{P}(|\vecp_\perp|,\gamma_w|\vecp_\perp|)$. The condition $H(\vecp,\veck)\geq 0$ leads to
\begin{align}
\label{eq:conditionH}
    \left[\gamma^2_w \vecp_\perp^2 +(k^z)^2\right]^2+ 4(\vecp_\perp\cdot \veck_\perp)^2+ 4\left[\gamma_w^2 \vecp^2_\perp+ (k^z)^2\right] \vecp_\perp\cdot \veck_\perp \geq 4\gamma_w^2 \vecp^2_\perp \left[ (k^z)^2+\veck_\perp^2+m^2_\phi\right]\,.
\end{align}
If $2\vecp_\perp \cdot \veck_\perp\gsim \gamma_w^2 \vecp_\perp^2+(k^z)^2$, one can show that
\begin{align}
    {\rm LHS} \lsim 16\vecp^2_\perp \veck^2_\perp \ll 4\gamma_w^2 \vecp^2_\perp \veck_\perp^2< {\rm RHS}\,,
\end{align}
so that Eq.~\eqref{eq:conditionH} cannot be satisfied. Therefore, we have 
\begin{align}
\label{eq:hierarchy-pk}
    2\vecp_\perp \cdot \veck_\perp \ll \gamma_w^2 \vecp_\perp^2+(k^z)^2\,.
\end{align}
As a consequence, one can ignore $2\vecp_\perp \cdot \veck_\perp$ in $H(\vecp,\veck)|_{p^z=\gamma_w |\vecp_\perp|}$,
\begin{align}
   H(\vecp,\veck)|_{p^z=|\vecp_\perp|} \rightarrow G(|\vecp_\perp|,|\veck_\perp|,k^z) \equiv \gamma_w^2 \vecp_\perp^2+(k^z)^2-2\gamma_w |\vecp_\perp| E^{(\phi)}_\veck\,.
\end{align}
And the constraint becomes
\begin{align}
\label{eq:constraint}
    G(|\vecp_\perp|,|\veck_\perp|,k^z)\geq 0 \quad \Leftrightarrow\quad  \gamma_w^2 \vecp_\perp^2+(k^z)^2 \geq 2\gamma_w |\vecp_\perp| E^{(\phi)}_\veck\,.
\end{align}
One can show that the above inequality automatically implies Eq.~\eqref{eq:hierarchy-pk}. 
From the constraint~\eqref{eq:constraint}, we can read the following information:
\begin{subequations}
\begin{align}
    &|\vecp_\perp|\geq \frac{2 m_\phi}{\gamma_w}\,,\\
    &0 \leq |\veck_\perp|\leq \sqrt{\frac{\gamma_w^2\vecp_\perp^2 }{4} -m_\phi^2} \,,\\
    \label{eq:kzmax}
    & 0\leq |k^z| \leq \sqrt{\gamma^2_w \vecp_\perp^2 -2\gamma_w |\vecp_\perp| \sqrt{\veck_\perp^2+m^2_\phi} }\equiv |k^z|_{\rm max}(|\vecp_\perp|,|\veck_\perp|)\,,
\end{align}
\end{subequations}
where we have defined $|k^z|_{\rm max}$.
The above defines the integral limits so that the integral~\eqref{eq:master-integral} is simplified to
\begin{align}
\label{eq:master-integral2}
    \P_{h\rightarrow 2\phi}= &\frac{\sqrt{2\pi T}\lambda^2 v_b^2 L_w^2}{2048\pi^3}\int_{2m_\phi/\gamma_w}^\infty \d |\vecp_\perp|\, |\vecp_\perp|^{1/2} \e^{-\frac{|\vecp_\perp|}{T}}\int_0^{\sqrt{\gamma_w^2\vecp_\perp^2/4 -m_\phi^2}}\d |\veck_\perp|\, |\veck_\perp|\notag\\
    &\qquad\qquad \times\int_{-|k^z|_{\rm max}(|\vecp_\perp|,|\veck_\perp|)}^{|k^z|_{\rm max}(|\vecp_\perp|,|\veck_\perp|)}\d k^z\, \frac{1}{E_\veck^{(\phi)}}\frac{\overline{\Delta p^z}}{\sqrt{G(|\vecp_\perp|,|\veck_\perp|,k^z)}} {\rm csch}^2 \left(\frac{\pi L_w \overline{\Delta p^z}}{2}\right)\,.
\end{align}
where the $z$-momentum transfer now reads
\begin{align}
    \overline{\Delta p^z}= \gamma_w |\vecp_\perp|-k^z-\sqrt{G(|\vecp_\perp|,|\veck_\perp|,k^z)}\,.
\end{align}

If one calculates the integral in Eq.~\eqref{eq:master-integral2} directly, there may be numerical instabilities.\footnote{This is the situation we encountered.} To avoid them, one can make use of the fact that the hyperbolic cosecant function ${\rm csch}(x)$ decreases fast with increasing $x$ and therefore add a cutoff for $x$:
\begin{align}
\label{eq:constraint2}
    \frac{\pi L_w \overline{\Delta p^z}}{2}\leq a\,,
\end{align}
where the cutoff $a$ characterizes the accuracy one aims for. For example, ${\rm csch}^2(10)\sim O(10^{-8})$ so that $a=10$ is good enough.

Now we work out how the above cutoff modifies the integral upper and lower limits. For fixed $|\vecp_\perp|$, the minimal of $\overline{\Delta p^z}$ is 
\begin{align}
(\overline{\Delta p^z})_{\rm min}(|\vecp_\perp|)=\gamma_w |\vecp_\perp| -\sqrt{\gamma_w^2 \vecp^2_\perp -4 m_\phi^2}\approx \frac{2m_\phi^2}{\gamma_w|\vecp_\perp| } \ {\rm at} \ \big\{k^z=\sqrt{\frac{\gamma_w^2\vecp_\perp^2}{4}-m^2_\phi},\ |\veck_\perp|=0\big\}\,.
\end{align}
Eq.~\eqref{eq:constraint2} immediately requires 
\begin{align}
\label{eq:pperpmin}
    (\overline{\Delta p^z})_{\rm min}(|\vecp_\perp|)\leq \frac{2a}{\pi L_w}\ \Rightarrow\ |\vecp_\perp|\geq \frac{\pi L_w m_\phi^2}{a\gamma_w}\equiv |\vecp_\perp|_{\rm min}\,,
\end{align}
where we have defined $|\vecp_\perp|_{\rm min}$. For fixed $|\vecp_\perp|$, $|\veck_\perp|$, the minimal of $\overline{\Delta p^z}$ is 
\begin{align}
    (\overline{\Delta p^z})_{\rm min}(|\vecp_\perp|,|\veck_\perp|)=\gamma_w |\vecp_\perp| -\sqrt{\gamma_w^2 \vecp^2_\perp -4(\veck_\perp^2+m_\phi^2)}  \ {\rm at} \ k^z=\sqrt{\frac{\gamma_w^2\vecp_\perp^2}{4}-(\veck_\perp^2+m^2_\phi)}\,.
\end{align}
Then Eq.~\eqref{eq:constraint2} requires 
\begin{align}
\label{eq:kperpmax}
    (\overline{\Delta p^z})_{\rm min}(|\vecp_\perp|,|\veck_\perp|) \leq \frac{2 a}{\pi L_w}\ 
     \Rightarrow\ \ |\veck_\perp| &\leq \frac{1}{\sqrt{2 L_w}}\sqrt{\gamma_w |\vecp_\perp|\left(\frac{2 a}{\pi}\right)-\frac{1}{2 L_w}\left(\frac{2 a}{\pi}\right)^2 - 2 L_w m_\phi^2 }\,,\notag\\
     &\approx
     \sqrt{\frac{ a\gamma_w |\vecp_\perp|}{\pi L_w} - m_\phi^2 }\equiv |\veck_\perp|_{\rm max}\,,
\end{align}
where we have defined $|\veck_\perp|_{\rm max}$.

\section{Numerical results and fit}
\label{sec:numerical-results}

In this section, we compute the integral ~\eqref{eq:master-integral2} numerically but using the integral upper and lower limits defined in Eqs.~\eqref{eq:kzmax},~\eqref{eq:pperpmin},~\eqref{eq:kperpmax}. Let
\begin{align}
    L_w =\frac{\kappa}{T}\,.
\end{align}
A typical value of $\kappa$ at
the electroweak transition with or without supersymmetry is of the order $\O(10)$~\cite{Moore:1995si,Moreno:1998bq}.

Defining the following dimensionless variables
\begin{align}
    x=\frac{|\vecp_\perp|}{T}\,,\qquad y=\frac{|\veck_\perp|}{T}\,,\qquad z=\frac{k^z}{T}\,,\qquad \xi=\frac{m_\phi}{T}\,,
\end{align}
the integral~\eqref{eq:master-integral2} can be written as
\begin{align}
    \label{eq:master-integral3}
    \P_{h\rightarrow 2\phi}&=\frac{\sqrt{2\pi} \lambda^2 v_b^2 \kappa^2 T^2}{2048\pi^3} \times \int_{x_{\rm min}}^\infty \d x\, x^{1/2} \e^{-x}\int_0^{y_{\rm max}(x)}\d y\, y \int_{-|z|_{\rm max}(x,y)}^{|z|_{\rm max}(x,y)}\d z\, \frac{\widetilde{\Delta p^z}\, {\rm csch}^2 \left(\frac{\kappa\pi \widetilde{\Delta p^z}}{2}\right)}{\widetilde{E}\widetilde{G}^{1/2}}\notag \\
    &\equiv B \times I(\xi,\gamma_w,\kappa)\,,
\end{align}
where 
\begin{align}
    \widetilde{E}=\sqrt{\xi^2+y^2+z^2}\,,\qquad \widetilde{G}=\gamma_w^2 x^2 +z^2 -2\gamma_w x \widetilde{E}\,,\qquad \widetilde{\Delta p^z}=\gamma_w x-z-\widetilde{G}^{1/2}
\end{align}
and
\begin{align}
    x_{\rm min}=\frac{\pi\kappa\xi^2}{a\gamma_w}\,,\qquad y_{\rm max}(x)=\sqrt{\frac{a \gamma_w x}{\pi\kappa}-\xi^2}\,,\qquad |z|_{\rm max}(x,y)=\sqrt{\gamma_w^2 x^2-2\gamma_w x \sqrt{\xi^2+y^2} }\,.
\end{align}

\subsection{\texorpdfstring{$L_w T=10$}{TEXT}}

We first consider $\kappa$ = 10. In the next subsection, we fit the $\kappa$-dependence in $I(\xi,\gamma_w,\kappa)$.

\begin{figure}[H]
    \centering
    \includegraphics[scale=0.5]{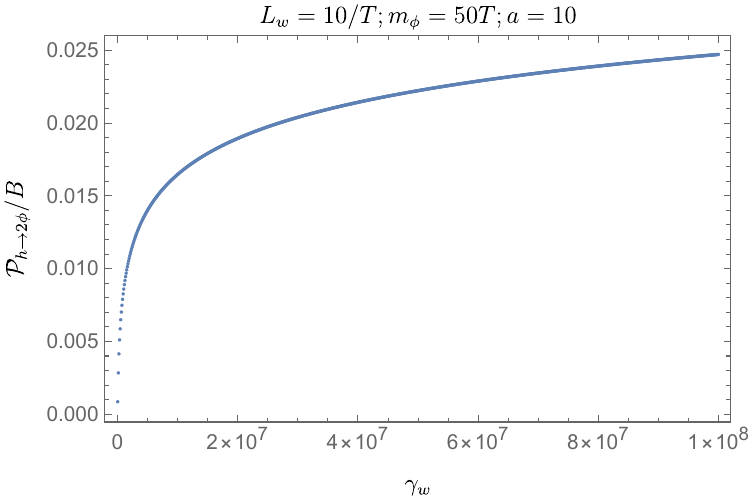}
    \hskip5pt
    \includegraphics[scale=0.5]{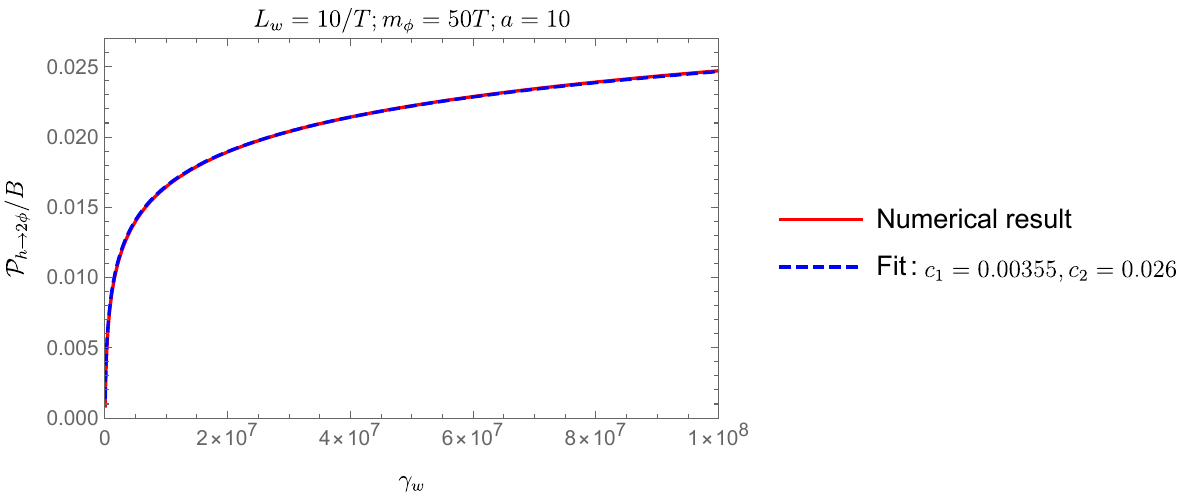}
    \caption{Left panel: numerical result of $\P_{h\rightarrow 2\phi}/B$ in the range $\gamma_w\in [5\times 10^4,10^8]$ for $m_\phi=50 T$. Right panel: comparison between the numerical result (red solid) and the fit (blue dashed) using the function~\eqref{eq:fit-function} with $c_1=0.00355, c_2=0.026$.}
    \label{fig:P1}
\end{figure}

\begin{figure}[H]
    \centering
    \includegraphics[scale=0.5]{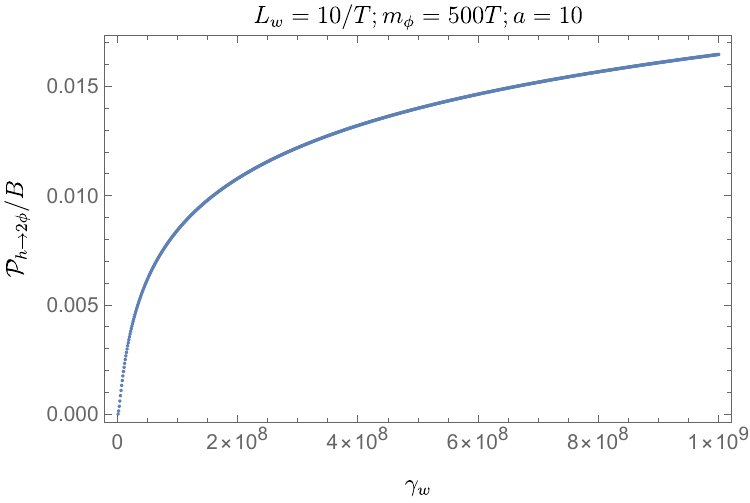}
    \hskip5pt
    \includegraphics[scale=0.5]{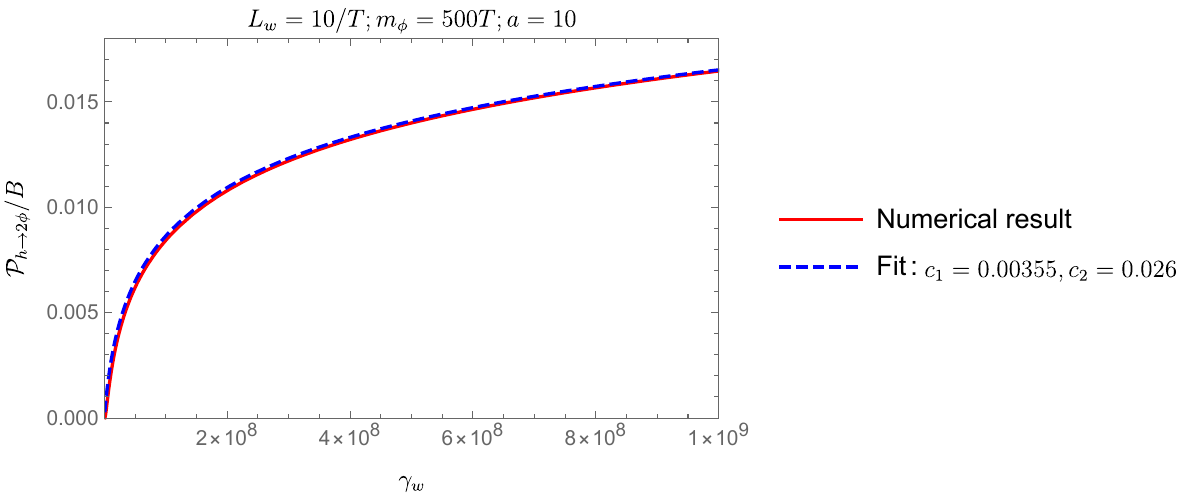}
    \caption{Left panel: numerical result of $\P_{h\rightarrow 2\phi}/B$ in the range $\gamma_w\in [10^6,10^9]$ for $m_\phi=500 T$. Right panel: comparison between the numerical result (red solid) and the fit (blue dashed) using the function~\eqref{eq:fit-function} with $c_1=0.00355, c_2=0.026$.}
    \label{fig:P2}
\end{figure}

\begin{figure}[H]
    \centering
    \includegraphics[scale=0.5]{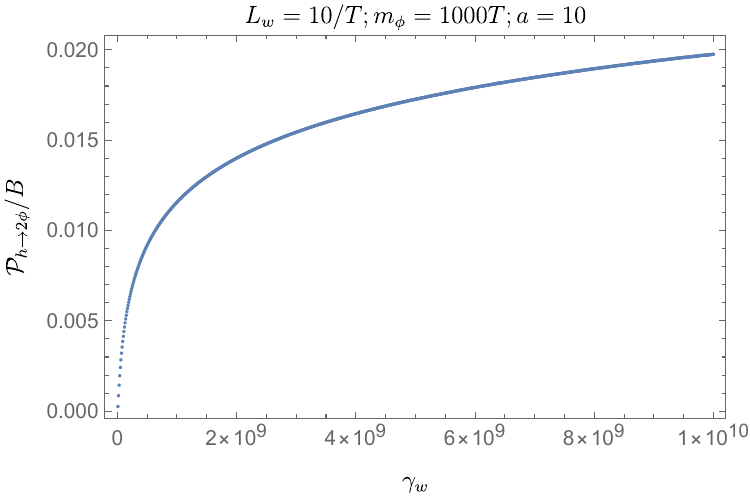}
    \hskip5pt
    \includegraphics[scale=0.5]{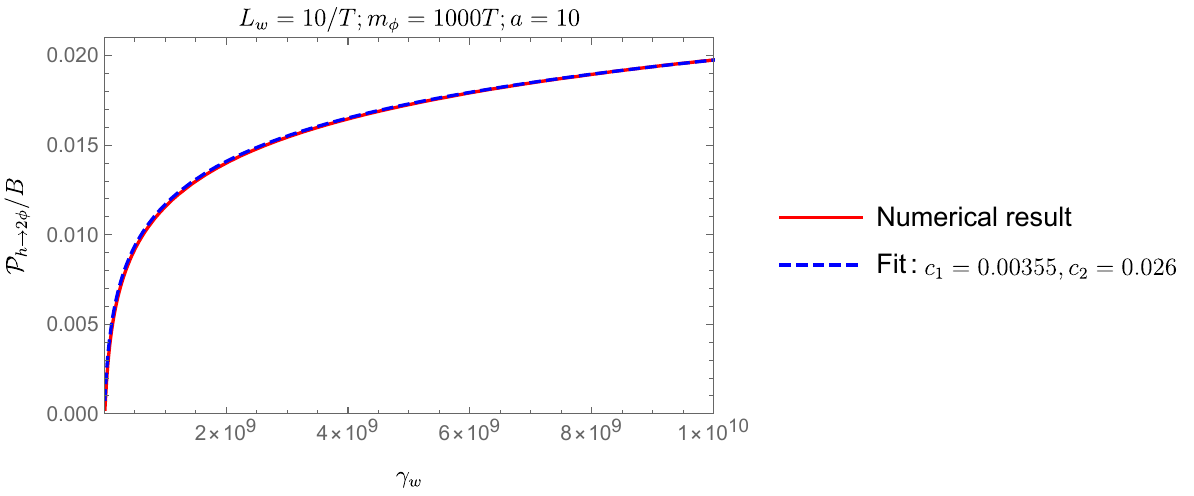}
    \caption{Left panel: numerical result of $\P_{h\rightarrow 2\phi}/B$ in the range $\gamma_w\in [10^7,10^{10}]$ for $m_\phi=1000 T$. Right panel: comparison between the numerical result (red solid) and the fit (blue dashed) using the function~\eqref{eq:fit-function} with $c_1=0.00355, c_2=0.026$.}
    \label{fig:P3}
\end{figure}

In the left panels of Figs.~\ref{fig:P1},~\ref{fig:P2},~\ref{fig:P3}, we show the numerical results of $\P_{h\rightarrow 2\phi}/B$ as a function of $\gamma_w$ for $m_\phi/T=50, 500, 1000$, respectively. Evidently, the backward pressure exhibits a discernible dependence on $\gamma_w$. The central question pertains to whether this dependence disappears as $\gamma_w\rightarrow\infty$. Such a trend could emerge if, for instance, the dependence follows a form like $\exp(-c_1\times \gamma_w^{-c_2})$ with $c_2>0$. To explore this, we undertake a fitting process for the numerical results using the following function:
\begin{align}
\label{eq:fit-function}
    f(\gamma_w)=c_1\log\left(1+c_2\frac{\gamma_w}{\xi^2}\right)\,.
\end{align}
The fitted curves with $c_1=0.00355, c_2=0.026$ are depicted in the right panels of Figs~\ref{fig:P1},\ref{fig:P2},\ref{fig:P3}. Notably, the dependence on $\gamma_w$ in the numerical results aligns remarkably well with the fit function.

\begin{figure}[H]
    \centering
    \includegraphics[scale=0.5]{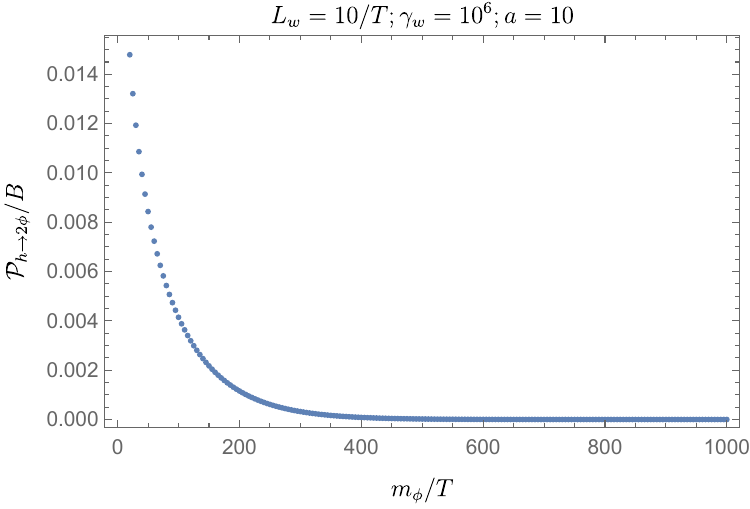}
    \hskip5pt
    \includegraphics[scale=0.5]{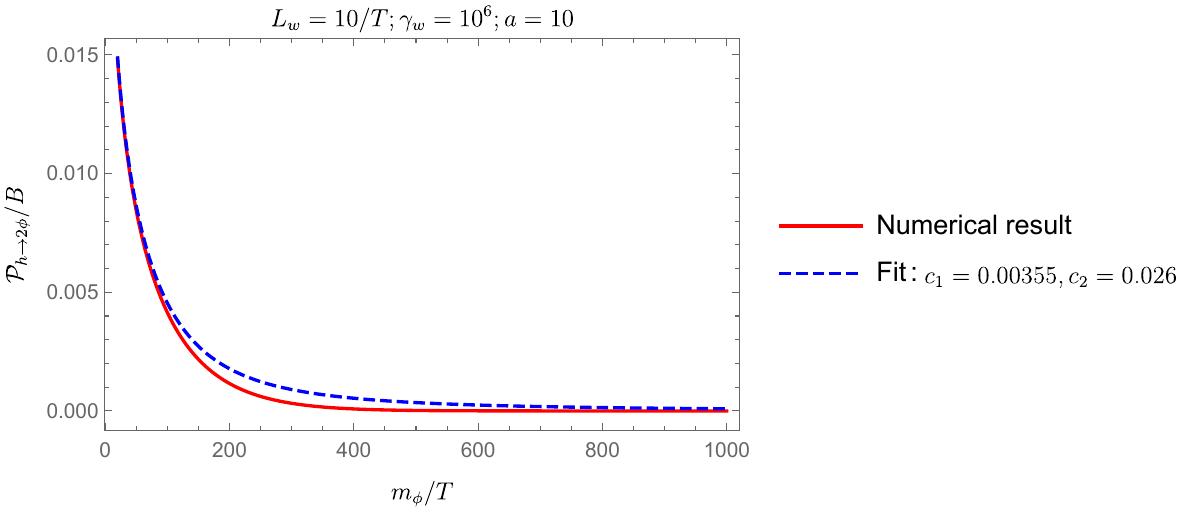}
    \caption{Left panel: numerical result of $\P_{h\rightarrow 2\phi}/B$ in the range $m_\phi/T\in [20,1000]$ for $\gamma_w=10^6$. Right panel: comparison between the numerical result (red solid) and the fit (blue dashed) using the function~\eqref{eq:fit-function} with $c_1=0.00355, c_2=0.026$.}
    \label{fig:P0}
\end{figure}

As a cross-check, one can also fix $\gamma_w$ and compare the numerical result with the fit function. In Fig.~\ref{fig:P0}, we show  $\P_{h\rightarrow 2\phi}/B$ as a function of $m_\phi/T$ for $\gamma_w=10^6$  as well as its comparison with the fit function with the same parameters $c_1$ and $c_2$ given above.

\subsection{Fit for the dependence on \texorpdfstring{$L_w T$}{TEXT}}

\begin{figure}[H]
    \centering
    \includegraphics[scale=0.54]{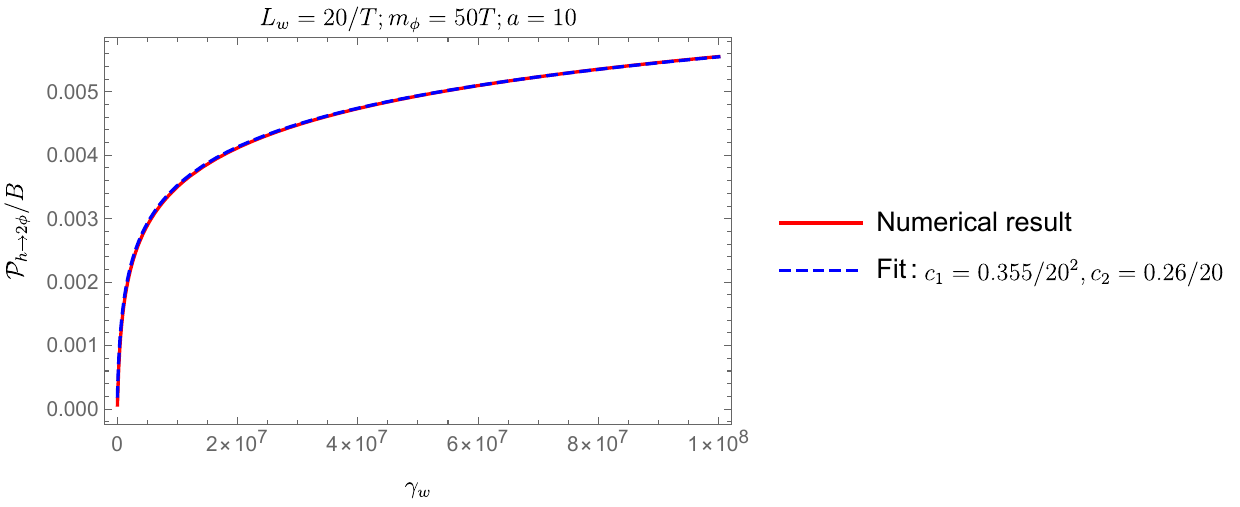}
    \includegraphics[scale=0.54]{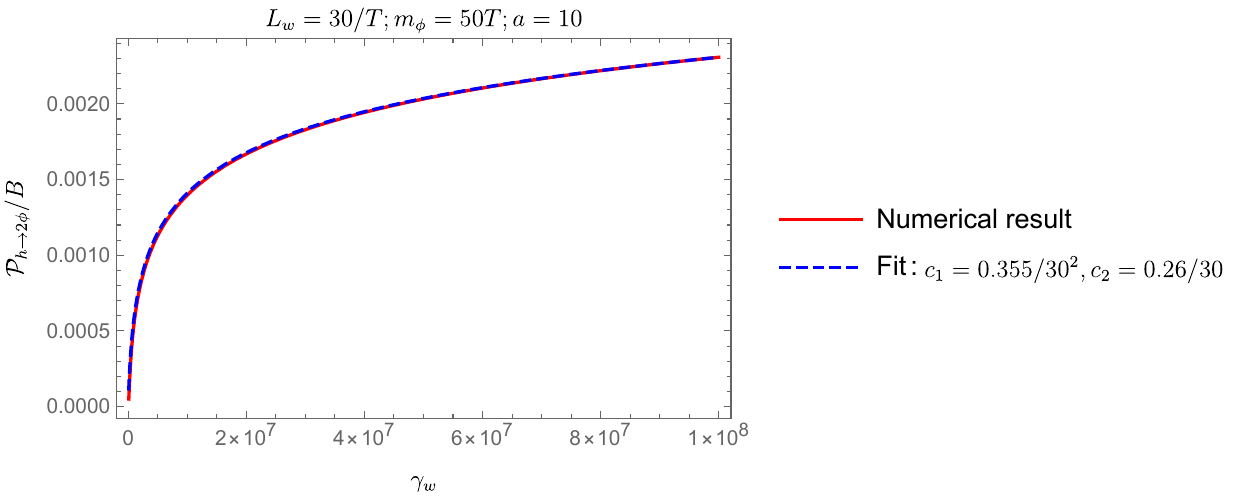}
    \includegraphics[scale=0.54]{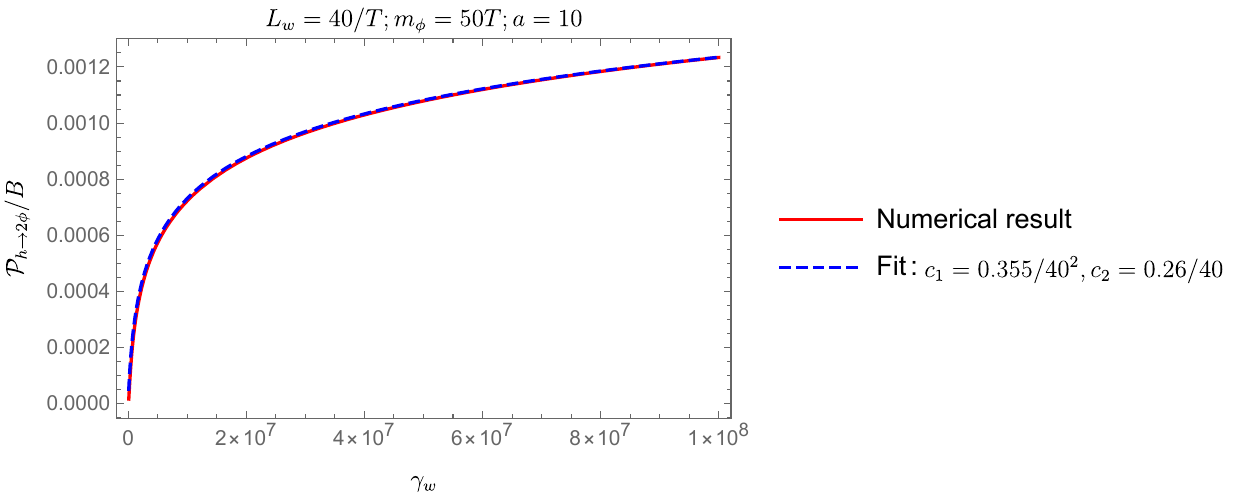}
    \includegraphics[scale=0.54]{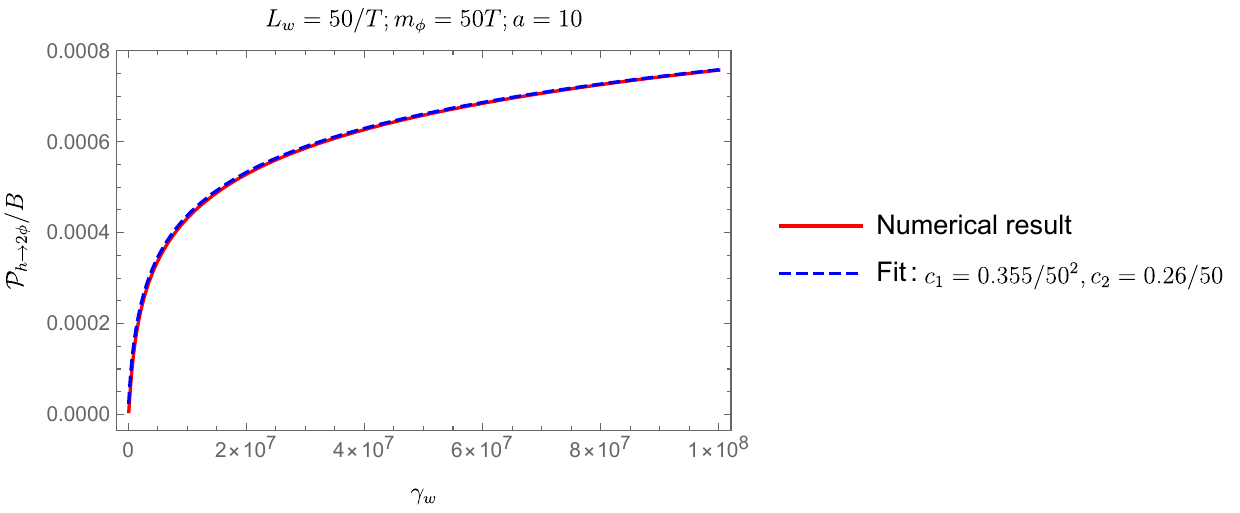}
    \caption{Comparison between the numerical results for different values of $L_w T$ and the fit using parameters given in Table~\ref{tab:fit-kappa}.}
    \label{fig:fit-kappa}
\end{figure}

Now we give the fit for the dependence of $I(\xi,\gamma_w,\kappa)$ on $\kappa$. The best fitted parameters $c_1$ and $c_2$ for $\kappa=10,...,50$ are given in Table~\ref{tab:fit-kappa}. We show some examples of the comparison between the numerical results and fit function in Fig.~\ref{fig:fit-kappa}. From the table, it is obvious that
\begin{align}
    c_1= 0.355\times \frac{1}{\kappa^2}\,,\quad c_2=0.26\times \frac{1}{\kappa}\,. 
\end{align}

\begin{table*}[ht]
\centering
    \begin{tabular}{|c|c|c|c|c|c|}
    \hline
    & $\kappa=10$   & $\kappa=20$ & $\kappa=30$ & $\kappa=40$ & 50 \\
    \hline         
   $c_1$ & $0.355/10^2$ & $0.355/20^2$ & $0.355/30^2$ & $0.355/40^2$ & $0.355/50^2$\\
    \hline
    $c_2$ & $0.26/10$ & $0.26/20$ &  $0.26/30$ & $0.26/40$ & $0.26/50$\\
    \hline
    \end{tabular}
    \caption{Fitted parameters of $c_1$ and $c_2$ for different values of $\kappa$.}
    \label{tab:fit-kappa}
\end{table*}

The final form of the pressure is fitted as
\begin{align}
\label{eq:central-result}
    \boxed{\P_{h\rightarrow 2\phi}\approx 1.4\times 10^{-5}\times\lambda^2 v_b^2 T^2\log\left(1+0.26\times \frac{\gamma_w T}{L_w m_\phi^2}\right)\,.}
\end{align}
To compare this friction with the leading-order friction due to the $1\rightarrow 1$ transmission of $\phi$-particles, we rewrite the above expression as
\begin{align}
    \P_{h\rightarrow 2\phi}=6.72\times 10^{-4}\lambda \log\left(1+0.26\times \frac{\gamma_w T}{L_w m_\phi^2}\right) \times \left(\frac{\Delta m_\phi^2 T^2}{24}\right)\,,
\end{align}
where $\Delta m_\phi^2=\lambda v_b^2/2$. This friction is typically much smaller than $\P_{\phi\rightarrow\phi}\simeq \Delta m_\phi^2 T^2/24$. 

\paragraph{Comment on potential sources of the discrepancy between our results and that in Ref.~\cite{Azatov:2020ufh}.}

As mentioned in the Introduction, the authors of Ref.~\cite{Azatov:2020ufh} found the frictional pressure $\P_{h\rightarrow 2\phi}$ to be independent of $\gamma_w$. Here, we just point out several approximations made in their analytic calculation which may or may not be the reason for the discrepancy between our numerical results and their analytic result.  

First, Ref.~\cite{Azatov:2020ufh} assumes $\vecp_\perp=0$ when calculating $\d \mathbb{P}_{h\rightarrow 2\phi}(\vecp)$. This itself is not a problem because one can always make $\vecp_\perp=0$ with a transverse boost. However, when substituting $\d \mathbb{P}_{h\rightarrow 2\phi}(\vecp)$ into $\P_{h\rightarrow 2\phi}$ (cf. Eq.~\eqref{eq:pressure}), one needs to take an inverse transverse boost to recover the $|\vecp_\perp|$-dependence in $\d \mathbb{P}_{h\rightarrow 2\phi}(\vecp)$ and then integrate over $|\vecp_\perp|$ in Eq.~\eqref{eq:pressure}. The second step is not done in Ref.~\cite{Azatov:2020ufh}.  

Second, Ref.~\cite{Azatov:2020ufh} assumes that for both $\phi$-particles, their energies are dominated by their $z$-momenta. Specifically, writing $k_1^z=\sqrt{(E_\vecp^{(h)})^2 x^2-\veck_{1,\perp}^2-m_\phi^2}$, $k_2^z=\sqrt{(E_\vecp^{(h)})^2 (1-x)^2-\veck_{2,\perp}^2-m_\phi^2}$, the authors of Ref.~\cite{Azatov:2020ufh} did the following expansion
\begin{align}
    k_1^z\approx E_\vecp^{(h)} x-\frac{1}{2}\frac{\veck_{1,\perp}^2+m_\phi^2}{ E_\vecp^{(h)} x}\,,\qquad  k_2^z\approx E_\veck^{(h)} (1-x)-\frac{1}{2}\frac{\veck_{2,\perp}^2+m_\phi^2}{ E_\vecp^{(h)} (1-x)}\,,
\end{align}
which is only true if 
\begin{align}
    (E_\vecp^{(h)})^2 x^2\gg \veck_{1,\perp}^2+m_\phi^2\,, \qquad (E_\vecp^{(h)})^2 (1-x)^2 \gg \veck_{2,\perp}^2+m_\phi^2\,.
\end{align}
As a consequence, the integral over the region with $(E_\vecp^{(h)})^2 x^2 \sim \veck_{1,\perp}^2+m_\phi^2$ or $(E_\vecp^{(h)})^2 (1-x)^2\sim \veck_{2,\perp}^2+m_\phi^2$ may not be captured precisely. 

Third, Ref.~\cite{Azatov:2020ufh} gives a stronger condition $L_w \Delta p^z<1$ than our Eq.~\eqref{eq:constraint2} with $a=10$. As a result, contributions from $\Delta p^z>1/L_w$ when the suppression from ${\rm csch}[\pi L_w\Delta p^z/2]$ is still not so large have been neglected. Further, our ${\rm csch}[x]$ is simply replaced with $1/x$ in their calculation which again could lead to some precision loss in the analytic result.

It is possible that the last two points mentioned might primarily result in a loss of precision, while the first point seems to introduce a slight consistency issue. Our approach, which involves initially integrating over $p^z$ using the method of steepest descent while retaining the $|\vecp_\perp|$-dependence in $\d\mathbb{P}_{h\rightarrow 2\phi}(\vecp)$, seems to offer better consistency. Our method actually provides a mathematical explanation for why $p^z\sim \gamma_w|\vecp_\perp|$. In fact, more careful treatment with the calculations in Ref.~\cite{Azatov:2020ufh} could recover the logarithmic dependence as well, in the same form (see Ref.~\cite{Baldes:2022oev}) but with some slightly different factors compared with our numerical fit~\eqref{eq:central-result}.\footnote{We thank Miguel Vanvlasselaer for making this comment.}

\subsection{The limit \texorpdfstring{$L_w\rightarrow 0$}{TEXT}}

In this subsection, we study the limit $L_w\rightarrow 0$. Usually, $L_w\gg 1/T$  is imposed to guarantee an adiabatic background, thus permitting the utilization of the WKB approximation for mode functions of free fields featuring a wall-dependent mass. In our specific context, the wall-dependent mass terms for both $\phi$ and $h$ can be safely neglected due to the condition $m_\phi \gg (T\sim v_b\sim m_h)$ (as discussed in Sec.~\ref{sec:review}). Consequently, it appears reasonable to explore the limit case of $L_w\rightarrow 0$ without necessitating a full rederivation.

One cannot take $L_w\rightarrow 0$ directly in Eq.~\eqref{eq:central-result}. To study this limit, we go back to Eq.~\eqref{eq:master-integral2}. For $L_w\rightarrow 0$, we have 
\begin{align}
    {\rm csch}^2\left(\frac{\pi L_w \overline{\Delta p^z}}{2}\right)\rightarrow \left(\frac{2}{\pi L_w \overline{\Delta p^z}}\right)^2\,.
\end{align}
Therefore, explicit factors of $L_w$ are canceled out in Eq.~\eqref{eq:master-integral2}. Using the same dimensionless variables, we have
\begin{align}
   \label{eq:master-integral3-Lw0}
    \P^{(L_w\rightarrow 0)}_{h\rightarrow 2\phi}&=\frac{\sqrt{2\pi} \lambda^2 v_b^2  T^2}{512\pi^5} \times \int_{x_{\rm min}}^\infty \d x\, x^{1/2} \e^{-x}\int_0^{y_{\rm max}(x)}\d y\, y \int_{-|z|_{\rm max}(x,y)}^{|z|_{\rm max}(x,y)}\d z\, \frac{1}{\widetilde{\Delta p^z}\widetilde{E}\widetilde{G}^{1/2}}\notag \\
    &\equiv C \times J(\xi,\gamma_w)\,, 
\end{align}
where $x_{\rm min}$, $y_{\rm max}(x)$ now are obtained from the corresponding integral lower and upper limits for $|\vecp_\perp|$ and $|\veck_\perp|$ in Eq.~\eqref{eq:master-integral2}, respectively.

\begin{figure}[H]
    \centering
    \includegraphics[scale=0.54]{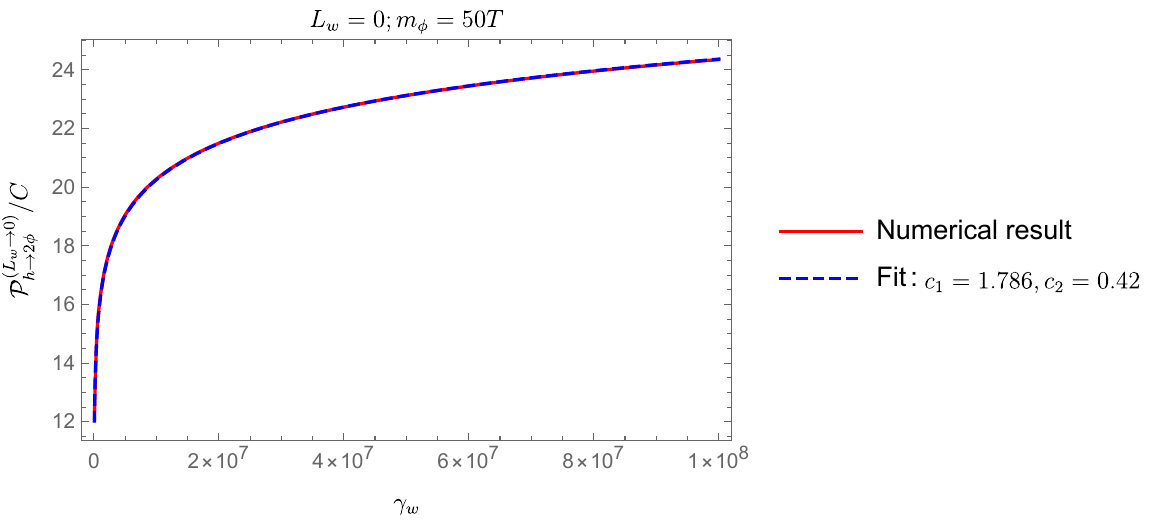}
    \includegraphics[scale=0.54]{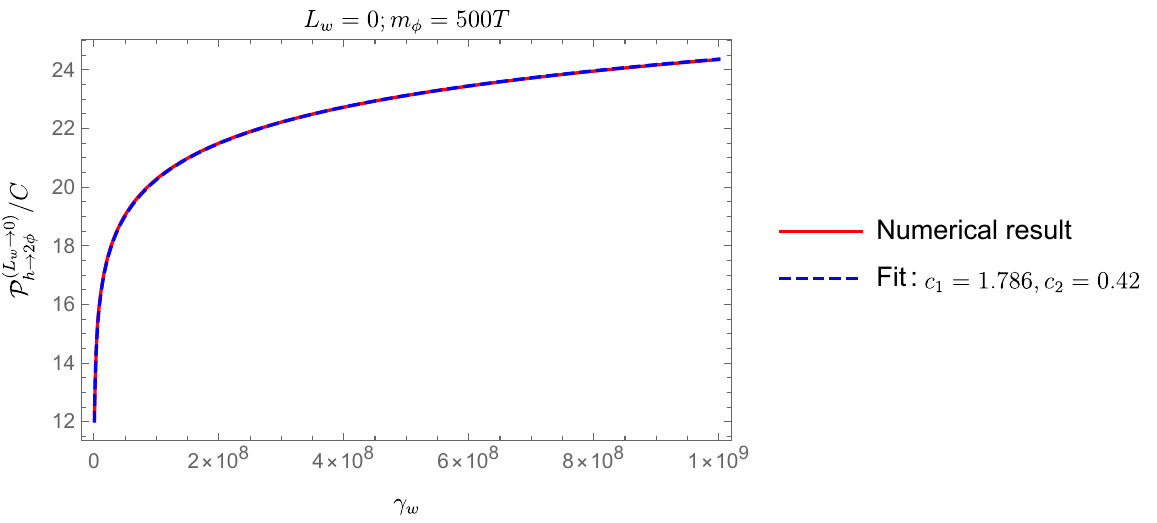}
    \includegraphics[scale=0.54]{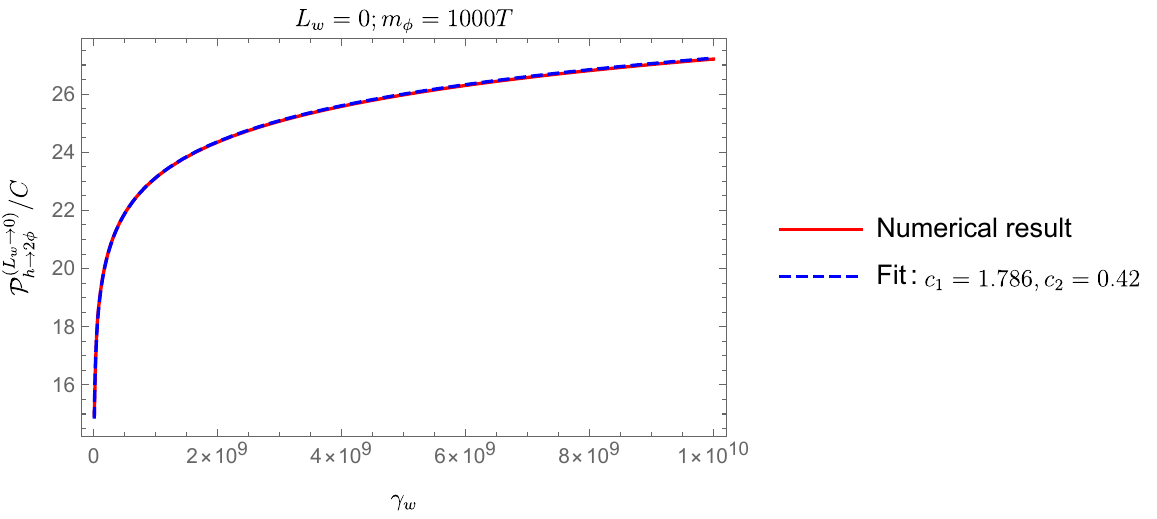}
    \caption{Comparison between the numerical results in the limit $L_w\rightarrow 0$ and the fit function~\eqref{eq:fit-function2} with $c_1=1.786$, $c_2=0.42$.}
    \label{fig:fit-Lw0}
\end{figure}

We fit the numerical results with the following function\footnote{We thank Aleksandr Azatov for suggesting this form of the pressure in the $L_w\rightarrow 0$ limit. Actually, the present subsection is motivated by his comment.}
\begin{align}
\label{eq:fit-function2}
    f^{(L_w\rightarrow 0)}(\gamma_w)=c_1\log\left(1+c_2\frac{\gamma_w}{\xi}\right)\,.
\end{align}
The comparison between the numerical results and the fit with $c_1=1.786, c_2=0.42$ for $m_\phi/T=50, 500, 1000$ is given in Fig.~\ref{fig:fit-Lw0}. We therefore conclude that in the limit $L_w\rightarrow 0$,
\begin{align}
    \P^{(L_w\rightarrow 0)}_{h\rightarrow 2\phi}\approx 2.86\times 10^{-5}\times \lambda^2 v_b^2 T^2 \log\left(1+ 0.42\times \frac{\gamma_w T}{m_\phi}\right)\,.
\end{align}


\section{Conclusion}
\label{sec:conclusion}

Understanding the friction exerted on the bubble wall is of paramount importance as it governs the wall's terminal velocity. Typically, this is a challenging endeavor necessitating the solution of nonequilibrium equations of motion for both the scalar field and plasma. However, for ultrarelativistic walls, it becomes possible to study the friction on a process-by-process basis. In this paper, we have undertaken a comprehensive reanalysis of the friction $\mathcal{P}_{h\rightarrow 2\phi}$ stemming from the $1\rightarrow 2$ light-to-heavy transition process ($h\rightarrow 2\phi$) within the ultrarelativistic regime. We have systematically streamlined the integral, making it amenable to numerical calculations. In our computation, we have accounted for finite-wall-width effects by introducing the Fourier transform of the bubble profile into the differential transition probability. Our numerical findings clearly indicate that the friction $\mathcal{P}_{h \rightarrow 2\phi}$ has a logarithmic dependence on $\gamma_w$. The numerical results can be perfectly fitted by Eq.~\eqref{eq:central-result}. The fit function indicates that $\P_{h\rightarrow 2\phi}$ is typically much smaller than $\P_{\phi\rightarrow \phi}$. The methodologies established in this study, with appropriate adaptations, hold the potential for investigating friction induced by other $1\rightarrow 2$ processes or dark matter particle production rates.

\section*{Acknowledgments}

I would like to express my gratitude to the host institute DESY, the organizers (Andreas Ekstedt and Jorinde van de Vis), and all the participants of the workshop “How fast does the bubble grow?”, during which the project was initiated. I am grateful to Miguel Vanvlasselaer for many discussions and helpful comments. I would also like to thank Aleksandr Azatov, Filippo Sala for helpful comments, Iason Baldes and Yann Gouttenoire for communication, and Zi-Liang Wang for the discussions on numerical computation. This work is supported by the UK Engineering and Physical Sciences Research Council (EPSRC), under Research Grant No. EP/V002821/1.

\newpage
\begin{appendix}
\renewcommand{\theequation}{\Alph{section}\arabic{equation}}

\end{appendix}

\newpage
\bibliographystyle{utphys}
\bibliography{ref}{}

\providecommand{\href}[2]{#2}\begingroup\raggedright\begin{thebibliography}{10}

\bibitem{Kuzmin:1985mm}
V.~A. Kuzmin, V.~A. Rubakov, and M.~E. Shaposhnikov, ``{On the Anomalous
  Electroweak Baryon Number Nonconservation in the Early Universe},''
  \href{http://dx.doi.org/10.1016/0370-2693(85)91028-7}{{\em Phys. Lett. B}
  {\bfseries 155} (1985) 36}.

\bibitem{Morrissey:2012db}
D.~E. Morrissey and M.~J. Ramsey-Musolf, ``{Electroweak baryogenesis},''
  \href{http://dx.doi.org/10.1088/1367-2630/14/12/125003}{{\em New J. Phys.}
  {\bfseries 14} (2012) 125003},
  \href{http://arxiv.org/abs/1206.2942}{{\ttfamily arXiv:1206.2942 [hep-ph]}}.

\bibitem{Garbrecht:2018mrp}
B.~Garbrecht, ``{Why is there more matter than antimatter? Calculational
  methods for leptogenesis and electroweak baryogenesis},''
  \href{http://dx.doi.org/10.1016/j.ppnp.2019.103727}{{\em Prog. Part. Nucl.
  Phys.} {\bfseries 110} (2020) 103727},
  \href{http://arxiv.org/abs/1812.02651}{{\ttfamily arXiv:1812.02651
  [hep-ph]}}.

\bibitem{Witten:1984rs}
E.~Witten, ``{Cosmic Separation of Phases},''
  \href{http://dx.doi.org/10.1103/PhysRevD.30.272}{{\em Phys. Rev. D}
  {\bfseries 30} (1984) 272--285}.

\bibitem{Kosowsky:1991ua}
A.~Kosowsky, M.~S. Turner, and R.~Watkins, ``{Gravitational radiation from
  colliding vacuum bubbles},''
  \href{http://dx.doi.org/10.1103/PhysRevD.45.4514}{{\em Phys. Rev. D}
  {\bfseries 45} (1992) 4514--4535}.

\bibitem{Kosowsky:1992vn}
A.~Kosowsky and M.~S. Turner, ``{Gravitational radiation from colliding vacuum
  bubbles: envelope approximation to many bubble collisions},''
  \href{http://dx.doi.org/10.1103/PhysRevD.47.4372}{{\em Phys. Rev. D}
  {\bfseries 47} (1993) 4372--4391},
  \href{http://arxiv.org/abs/astro-ph/9211004}{{\ttfamily
  arXiv:astro-ph/9211004}}.

\bibitem{Kamionkowski:1993fg}
M.~Kamionkowski, A.~Kosowsky, and M.~S. Turner, ``{Gravitational radiation from
  first order phase transitions},''
  \href{http://dx.doi.org/10.1103/PhysRevD.49.2837}{{\em Phys. Rev. D}
  {\bfseries 49} (1994) 2837--2851},
  \href{http://arxiv.org/abs/astro-ph/9310044}{{\ttfamily
  arXiv:astro-ph/9310044}}.

\bibitem{Huber:2008hg}
S.~J. Huber and T.~Konstandin, ``{Gravitational Wave Production by Collisions:
  More Bubbles},'' \href{http://dx.doi.org/10.1088/1475-7516/2008/09/022}{{\em
  JCAP} {\bfseries 09} (2008) 022},
  \href{http://arxiv.org/abs/0806.1828}{{\ttfamily arXiv:0806.1828 [hep-ph]}}.

\bibitem{Hindmarsh:2013xza}
M.~Hindmarsh, S.~J. Huber, K.~Rummukainen, and D.~J. Weir, ``{Gravitational
  waves from the sound of a first order phase transition},''
  \href{http://dx.doi.org/10.1103/PhysRevLett.112.041301}{{\em Phys. Rev.
  Lett.} {\bfseries 112} (2014) 041301},
  \href{http://arxiv.org/abs/1304.2433}{{\ttfamily arXiv:1304.2433 [hep-ph]}}.

\bibitem{Grojean:2006bp}
C.~Grojean and G.~Servant, ``{Gravitational Waves from Phase Transitions at the
  Electroweak Scale and Beyond},''
  \href{http://dx.doi.org/10.1103/PhysRevD.75.043507}{{\em Phys. Rev. D}
  {\bfseries 75} (2007) 043507},
  \href{http://arxiv.org/abs/hep-ph/0607107}{{\ttfamily arXiv:hep-ph/0607107}}.

\bibitem{Caprini:2018mtu}
C.~Caprini and D.~G. Figueroa, ``{Cosmological Backgrounds of Gravitational
  Waves},'' \href{http://dx.doi.org/10.1088/1361-6382/aac608}{{\em Class.
  Quant. Grav.} {\bfseries 35} no.~16, (2018) 163001},
  \href{http://arxiv.org/abs/1801.04268}{{\ttfamily arXiv:1801.04268
  [astro-ph.CO]}}.

\bibitem{Mazumdar:2018dfl}
A.~Mazumdar and G.~White, ``{Review of cosmic phase transitions: their
  significance and experimental signatures},''
  \href{http://dx.doi.org/10.1088/1361-6633/ab1f55}{{\em Rept. Prog. Phys.}
  {\bfseries 82} no.~7, (2019) 076901},
  \href{http://arxiv.org/abs/1811.01948}{{\ttfamily arXiv:1811.01948
  [hep-ph]}}.

\bibitem{Caprini:2019egz}
C.~Caprini {\em et~al.}, ``{Detecting gravitational waves from cosmological
  phase transitions with LISA: an update},''
  \href{http://dx.doi.org/10.1088/1475-7516/2020/03/024}{{\em JCAP} {\bfseries
  03} (2020) 024}, \href{http://arxiv.org/abs/1910.13125}{{\ttfamily
  arXiv:1910.13125 [astro-ph.CO]}}.

\bibitem{LISACosmologyWorkingGroup:2022jok}
{\bfseries LISA Cosmology Working Group} Collaboration, P.~Auclair {\em
  et~al.}, ``{Cosmology with the Laser Interferometer Space Antenna},''
  \href{http://arxiv.org/abs/2204.05434}{{\ttfamily arXiv:2204.05434
  [astro-ph.CO]}}.

\bibitem{Badger:2022nwo}
C.~Badger {\em et~al.}, ``{Probing early Universe supercooled phase transitions
  with gravitational wave data},''
  \href{http://dx.doi.org/10.1103/PhysRevD.107.023511}{{\em Phys. Rev. D}
  {\bfseries 107} no.~2, (2023) 023511},
  \href{http://arxiv.org/abs/2209.14707}{{\ttfamily arXiv:2209.14707
  [hep-ph]}}.

\bibitem{Athron:2023xlk}
P.~Athron, C.~Bal\'azs, A.~Fowlie, L.~Morris, and L.~Wu, ``{Cosmological phase
  transitions: from perturbative particle physics to gravitational waves},''
  \href{http://arxiv.org/abs/2305.02357}{{\ttfamily arXiv:2305.02357
  [hep-ph]}}.

\bibitem{NANOGrav:2023gor}
{\bfseries NANOGrav} Collaboration, G.~Agazie {\em et~al.}, ``{The NANOGrav 15
  yr Data Set: Evidence for a Gravitational-wave Background},''
  \href{http://dx.doi.org/10.3847/2041-8213/acdac6}{{\em Astrophys. J. Lett.}
  {\bfseries 951} no.~1, (2023) L8},
  \href{http://arxiv.org/abs/2306.16213}{{\ttfamily arXiv:2306.16213
  [astro-ph.HE]}}.

\bibitem{EPTA:2023fyk}
{\bfseries EPTA} Collaboration, J.~Antoniadis {\em et~al.}, ``{The second data
  release from the European Pulsar Timing Array III. Search for gravitational
  wave signals},'' \href{http://arxiv.org/abs/2306.16214}{{\ttfamily
  arXiv:2306.16214 [astro-ph.HE]}}.

\bibitem{Reardon:2023gzh}
D.~J. Reardon {\em et~al.}, ``{Search for an Isotropic Gravitational-wave
  Background with the Parkes Pulsar Timing Array},''
  \href{http://dx.doi.org/10.3847/2041-8213/acdd02}{{\em Astrophys. J. Lett.}
  {\bfseries 951} no.~1, (2023) L6},
  \href{http://arxiv.org/abs/2306.16215}{{\ttfamily arXiv:2306.16215
  [astro-ph.HE]}}.

\bibitem{Xu:2023wog}
H.~Xu {\em et~al.}, ``{Searching for the Nano-Hertz Stochastic Gravitational
  Wave Background with the Chinese Pulsar Timing Array Data Release I},''
  \href{http://dx.doi.org/10.1088/1674-4527/acdfa5}{{\em Res. Astron.
  Astrophys.} {\bfseries 23} no.~7, (2023) 075024},
  \href{http://arxiv.org/abs/2306.16216}{{\ttfamily arXiv:2306.16216
  [astro-ph.HE]}}.

\bibitem{Cline:2020jre}
J.~M. Cline and K.~Kainulainen, ``{Electroweak baryogenesis at high bubble wall
  velocities},'' \href{http://dx.doi.org/10.1103/PhysRevD.101.063525}{{\em
  Phys. Rev. D} {\bfseries 101} no.~6, (2020) 063525},
  \href{http://arxiv.org/abs/2001.00568}{{\ttfamily arXiv:2001.00568
  [hep-ph]}}.

\bibitem{Cline:2021dkf}
J.~M. Cline and B.~Laurent, ``{Electroweak baryogenesis from light fermion
  sources: A critical study},''
  \href{http://dx.doi.org/10.1103/PhysRevD.104.083507}{{\em Phys. Rev. D}
  {\bfseries 104} no.~8, (2021) 083507},
  \href{http://arxiv.org/abs/2108.04249}{{\ttfamily arXiv:2108.04249
  [hep-ph]}}.

\bibitem{Ellis:2022lft}
J.~Ellis, M.~Lewicki, M.~Merchand, J.~M. No, and M.~Zych, ``{The scalar singlet
  extension of the Standard Model: gravitational waves versus baryogenesis},''
  \href{http://dx.doi.org/10.1007/JHEP01(2023)093}{{\em JHEP} {\bfseries 01}
  (2023) 093}, \href{http://arxiv.org/abs/2210.16305}{{\ttfamily
  arXiv:2210.16305 [hep-ph]}}.

\bibitem{Caprini:2015zlo}
C.~Caprini {\em et~al.}, ``{Science with the space-based interferometer eLISA.
  II: Gravitational waves from cosmological phase transitions},''
  \href{http://dx.doi.org/10.1088/1475-7516/2016/04/001}{{\em JCAP} {\bfseries
  04} (2016) 001}, \href{http://arxiv.org/abs/1512.06239}{{\ttfamily
  arXiv:1512.06239 [astro-ph.CO]}}.

\bibitem{Gowling:2021gcy}
C.~Gowling and M.~Hindmarsh, ``{Observational prospects for phase transitions
  at LISA: Fisher matrix analysis},''
  \href{http://dx.doi.org/10.1088/1475-7516/2021/10/039}{{\em JCAP} {\bfseries
  10} (2021) 039}, \href{http://arxiv.org/abs/2106.05984}{{\ttfamily
  arXiv:2106.05984 [astro-ph.CO]}}.

\bibitem{Liu:1992tn}
B.-H. Liu, L.~D. McLerran, and N.~Turok, ``{Bubble nucleation and growth at a
  baryon number producing electroweak phase transition},''
  \href{http://dx.doi.org/10.1103/PhysRevD.46.2668}{{\em Phys. Rev. D}
  {\bfseries 46} (1992) 2668--2688}.

\bibitem{Moore:1995ua}
G.~D. Moore and T.~Prokopec, ``{Bubble wall velocity in a first order
  electroweak phase transition},''
  \href{http://dx.doi.org/10.1103/PhysRevLett.75.777}{{\em Phys. Rev. Lett.}
  {\bfseries 75} (1995) 777--780},
  \href{http://arxiv.org/abs/hep-ph/9503296}{{\ttfamily arXiv:hep-ph/9503296}}.

\bibitem{Moore:1995si}
G.~D. Moore and T.~Prokopec, ``{How fast can the wall move? A Study of the
  electroweak phase transition dynamics},''
  \href{http://dx.doi.org/10.1103/PhysRevD.52.7182}{{\em Phys. Rev. D}
  {\bfseries 52} (1995) 7182--7204},
  \href{http://arxiv.org/abs/hep-ph/9506475}{{\ttfamily arXiv:hep-ph/9506475}}.

\bibitem{Konstandin:2014zta}
T.~Konstandin, G.~Nardini, and I.~Rues, ``{From Boltzmann equations to steady
  wall velocities},''
  \href{http://dx.doi.org/10.1088/1475-7516/2014/09/028}{{\em JCAP} {\bfseries
  09} (2014) 028}, \href{http://arxiv.org/abs/1407.3132}{{\ttfamily
  arXiv:1407.3132 [hep-ph]}}.

\bibitem{Ignatius:1993qn}
J.~Ignatius, K.~Kajantie, H.~Kurki-Suonio, and M.~Laine, ``{The growth of
  bubbles in cosmological phase transitions},''
  \href{http://dx.doi.org/10.1103/PhysRevD.49.3854}{{\em Phys. Rev. D}
  {\bfseries 49} (1994) 3854--3868},
  \href{http://arxiv.org/abs/astro-ph/9309059}{{\ttfamily
  arXiv:astro-ph/9309059}}.

\bibitem{Heckler:1994uu}
A.~F. Heckler, ``{The Effects of electroweak phase transition dynamics on
  baryogenesis and primordial nucleosynthesis},''
  \href{http://dx.doi.org/10.1103/PhysRevD.51.405}{{\em Phys. Rev. D}
  {\bfseries 51} (1995) 405--428},
  \href{http://arxiv.org/abs/astro-ph/9407064}{{\ttfamily
  arXiv:astro-ph/9407064}}.

\bibitem{Kurki-Suonio:1996gkq}
H.~Kurki-Suonio and M.~Laine, ``{Real time history of the cosmological
  electroweak phase transition},''
  \href{http://dx.doi.org/10.1103/PhysRevLett.77.3951}{{\em Phys. Rev. Lett.}
  {\bfseries 77} (1996) 3951--3954},
  \href{http://arxiv.org/abs/hep-ph/9607382}{{\ttfamily arXiv:hep-ph/9607382}}.

\bibitem{Megevand:2009ut}
A.~Megevand and A.~D. Sanchez, ``{Detonations and deflagrations in cosmological
  phase transitions},''
  \href{http://dx.doi.org/10.1016/j.nuclphysb.2009.05.007}{{\em Nucl. Phys. B}
  {\bfseries 820} (2009) 47--74},
  \href{http://arxiv.org/abs/0904.1753}{{\ttfamily arXiv:0904.1753 [hep-ph]}}.

\bibitem{Megevand:2009gh}
A.~Megevand and A.~D. Sanchez, ``{Velocity of electroweak bubble walls},''
  \href{http://dx.doi.org/10.1016/j.nuclphysb.2009.09.019}{{\em Nucl. Phys. B}
  {\bfseries 825} (2010) 151--176},
  \href{http://arxiv.org/abs/0908.3663}{{\ttfamily arXiv:0908.3663 [hep-ph]}}.

\bibitem{Espinosa:2010hh}
J.~R. Espinosa, T.~Konstandin, J.~M. No, and G.~Servant, ``{Energy Budget of
  Cosmological First-order Phase Transitions},''
  \href{http://dx.doi.org/10.1088/1475-7516/2010/06/028}{{\em JCAP} {\bfseries
  06} (2010) 028}, \href{http://arxiv.org/abs/1004.4187}{{\ttfamily
  arXiv:1004.4187 [hep-ph]}}.

\bibitem{Moore:2000wx}
G.~D. Moore, ``{Electroweak bubble wall friction: Analytic results},''
  \href{http://dx.doi.org/10.1088/1126-6708/2000/03/006}{{\em JHEP} {\bfseries
  03} (2000) 006}, \href{http://arxiv.org/abs/hep-ph/0001274}{{\ttfamily
  arXiv:hep-ph/0001274}}.

\bibitem{John:2000zq}
P.~John and M.~G. Schmidt, ``{Do stops slow down electroweak bubble walls?},''
  \href{http://dx.doi.org/10.1016/S0550-3213(00)00768-9}{{\em Nucl. Phys. B}
  {\bfseries 598} (2001) 291--305},
  \href{http://arxiv.org/abs/hep-ph/0002050}{{\ttfamily arXiv:hep-ph/0002050}}.
  [Erratum: Nucl.Phys.B 648, 449--452 (2003)].

\bibitem{Huber:2011aa}
S.~J. Huber and M.~Sopena, ``{The bubble wall velocity in the minimal
  supersymmetric light stop scenario},''
  \href{http://dx.doi.org/10.1103/PhysRevD.85.103507}{{\em Phys. Rev. D}
  {\bfseries 85} (2012) 103507},
  \href{http://arxiv.org/abs/1112.1888}{{\ttfamily arXiv:1112.1888 [hep-ph]}}.

\bibitem{Huber:2013kj}
S.~J. Huber and M.~Sopena, ``{An efficient approach to electroweak bubble
  velocities},'' \href{http://arxiv.org/abs/1302.1044}{{\ttfamily
  arXiv:1302.1044 [hep-ph]}}.

\bibitem{Dorsch:2018pat}
G.~C. Dorsch, S.~J. Huber, and T.~Konstandin, ``{Bubble wall velocities in the
  Standard Model and beyond},''
  \href{http://dx.doi.org/10.1088/1475-7516/2018/12/034}{{\em JCAP} {\bfseries
  12} (2018) 034}, \href{http://arxiv.org/abs/1809.04907}{{\ttfamily
  arXiv:1809.04907 [hep-ph]}}.

\bibitem{Wang:2020zlf}
X.~Wang, F.~P. Huang, and X.~Zhang, ``{Bubble wall velocity beyond leading-log
  approximation in electroweak phase transition},''
  \href{http://arxiv.org/abs/2011.12903}{{\ttfamily arXiv:2011.12903
  [hep-ph]}}.

\bibitem{Laurent:2022jrs}
B.~Laurent and J.~M. Cline, ``{First principles determination of bubble wall
  velocity},'' \href{http://dx.doi.org/10.1103/PhysRevD.106.023501}{{\em Phys.
  Rev. D} {\bfseries 106} no.~2, (2022) 023501},
  \href{http://arxiv.org/abs/2204.13120}{{\ttfamily arXiv:2204.13120
  [hep-ph]}}.

\bibitem{Jiang:2022btc}
S.~Jiang, F.~P. Huang, and X.~Wang, ``{Bubble wall velocity during electroweak
  phase transition in the inert doublet model},''
  \href{http://arxiv.org/abs/2211.13142}{{\ttfamily arXiv:2211.13142
  [hep-ph]}}.

\bibitem{BarrosoMancha:2020fay}
M.~Barroso~Mancha, T.~Prokopec, and B.~Swiezewska, ``{Field-theoretic
  derivation of bubble-wall force},''
  \href{http://dx.doi.org/10.1007/JHEP01(2021)070}{{\em JHEP} {\bfseries 01}
  (2021) 070}, \href{http://arxiv.org/abs/2005.10875}{{\ttfamily
  arXiv:2005.10875 [hep-th]}}.

\bibitem{Balaji:2020yrx}
S.~Balaji, M.~Spannowsky, and C.~Tamarit, ``{Cosmological bubble friction in
  local equilibrium},''
  \href{http://dx.doi.org/10.1088/1475-7516/2021/03/051}{{\em JCAP} {\bfseries
  03} (2021) 051}, \href{http://arxiv.org/abs/2010.08013}{{\ttfamily
  arXiv:2010.08013 [hep-ph]}}.

\bibitem{Ai:2021kak}
W.-Y. Ai, B.~Garbrecht, and C.~Tamarit, ``{Bubble wall velocities in local
  equilibrium},'' \href{http://dx.doi.org/10.1088/1475-7516/2022/03/015}{{\em
  JCAP} {\bfseries 03} no.~03, (2022) 015},
  \href{http://arxiv.org/abs/2109.13710}{{\ttfamily arXiv:2109.13710
  [hep-ph]}}.

\bibitem{Ai:2023see}
W.-Y. Ai, B.~Laurent, and J.~van~de Vis, ``{Model-independent bubble wall
  velocities in local thermal equilibrium},''
  \href{http://dx.doi.org/10.1088/1475-7516/2023/07/002}{{\em JCAP} {\bfseries
  07} (2023) 002}, \href{http://arxiv.org/abs/2303.10171}{{\ttfamily
  arXiv:2303.10171 [astro-ph.CO]}}.

\bibitem{Wang:2022txy}
S.-J. Wang and Z.-Y. Yuwen, ``{Hydrodynamic backreaction force of cosmological
  bubble expansion},''
  \href{http://dx.doi.org/10.1103/PhysRevD.107.023501}{{\em Phys. Rev. D}
  {\bfseries 107} no.~2, (2023) 023501},
  \href{http://arxiv.org/abs/2205.02492}{{\ttfamily arXiv:2205.02492
  [hep-ph]}}.

\bibitem{Konstandin:2010dm}
T.~Konstandin and J.~M. No, ``{Hydrodynamic obstruction to bubble expansion},''
  \href{http://dx.doi.org/10.1088/1475-7516/2011/02/008}{{\em JCAP} {\bfseries
  02} (2011) 008}, \href{http://arxiv.org/abs/1011.3735}{{\ttfamily
  arXiv:1011.3735 [hep-ph]}}.

\bibitem{Bodeker:2009qy}
D.~Bodeker and G.~D. Moore, ``{Can electroweak bubble walls run away?},''
  \href{http://dx.doi.org/10.1088/1475-7516/2009/05/009}{{\em JCAP} {\bfseries
  05} (2009) 009}, \href{http://arxiv.org/abs/0903.4099}{{\ttfamily
  arXiv:0903.4099 [hep-ph]}}.

\bibitem{Bodeker:2017cim}
D.~Bodeker and G.~D. Moore, ``{Electroweak Bubble Wall Speed Limit},''
  \href{http://dx.doi.org/10.1088/1475-7516/2017/05/025}{{\em JCAP} {\bfseries
  05} (2017) 025}, \href{http://arxiv.org/abs/1703.08215}{{\ttfamily
  arXiv:1703.08215 [hep-ph]}}.

\bibitem{Hoche:2020ysm}
S.~H\"oche, J.~Kozaczuk, A.~J. Long, J.~Turner, and Y.~Wang, ``{Towards an
  all-orders calculation of the electroweak bubble wall velocity},''
  \href{http://dx.doi.org/10.1088/1475-7516/2021/03/009}{{\em JCAP} {\bfseries
  03} (2021) 009}, \href{http://arxiv.org/abs/2007.10343}{{\ttfamily
  arXiv:2007.10343 [hep-ph]}}.

\bibitem{Azatov:2020ufh}
A.~Azatov and M.~Vanvlasselaer, ``{Bubble wall velocity: heavy physics
  effects},'' \href{http://dx.doi.org/10.1088/1475-7516/2021/01/058}{{\em JCAP}
  {\bfseries 01} (2021) 058}, \href{http://arxiv.org/abs/2010.02590}{{\ttfamily
  arXiv:2010.02590 [hep-ph]}}.

\bibitem{Gouttenoire:2021kjv}
Y.~Gouttenoire, R.~Jinno, and F.~Sala, ``{Friction pressure on relativistic
  bubble walls},'' \href{http://dx.doi.org/10.1007/JHEP05(2022)004}{{\em JHEP}
  {\bfseries 05} (2022) 004}, \href{http://arxiv.org/abs/2112.07686}{{\ttfamily
  arXiv:2112.07686 [hep-ph]}}.

\bibitem{Dine:1992wr}
M.~Dine, R.~G. Leigh, P.~Y. Huet, A.~D. Linde, and D.~A. Linde, ``{Towards the
  theory of the electroweak phase transition},''
  \href{http://dx.doi.org/10.1103/PhysRevD.46.550}{{\em Phys. Rev. D}
  {\bfseries 46} (1992) 550--571},
  \href{http://arxiv.org/abs/hep-ph/9203203}{{\ttfamily arXiv:hep-ph/9203203}}.

\bibitem{Azatov:2021ifm}
A.~Azatov, M.~Vanvlasselaer, and W.~Yin, ``{Dark Matter production from
  relativistic bubble walls},''
  \href{http://dx.doi.org/10.1007/JHEP03(2021)288}{{\em JHEP} {\bfseries 03}
  (2021) 288}, \href{http://arxiv.org/abs/2101.05721}{{\ttfamily
  arXiv:2101.05721 [hep-ph]}}.

\bibitem{Azatov:2021irb}
A.~Azatov, M.~Vanvlasselaer, and W.~Yin, ``{Baryogenesis via relativistic
  bubble walls},'' \href{http://dx.doi.org/10.1007/JHEP10(2021)043}{{\em JHEP}
  {\bfseries 10} (2021) 043}, \href{http://arxiv.org/abs/2106.14913}{{\ttfamily
  arXiv:2106.14913 [hep-ph]}}.

\bibitem{Baldes:2021vyz}
I.~Baldes, S.~Blasi, A.~Mariotti, A.~Sevrin, and K.~Turbang, ``{Baryogenesis
  via relativistic bubble expansion},''
  \href{http://dx.doi.org/10.1103/PhysRevD.104.115029}{{\em Phys. Rev. D}
  {\bfseries 104} no.~11, (2021) 115029},
  \href{http://arxiv.org/abs/2106.15602}{{\ttfamily arXiv:2106.15602
  [hep-ph]}}.

\bibitem{Huang:2022vkf}
P.~Huang and K.-P. Xie, ``{Leptogenesis triggered by a first-order phase
  transition},'' \href{http://dx.doi.org/10.1007/JHEP09(2022)052}{{\em JHEP}
  {\bfseries 09} (2022) 052}, \href{http://arxiv.org/abs/2206.04691}{{\ttfamily
  arXiv:2206.04691 [hep-ph]}}.

\bibitem{Azatov:2022tii}
A.~Azatov, G.~Barni, S.~Chakraborty, M.~Vanvlasselaer, and W.~Yin,
  ``{Ultra-relativistic bubbles from the simplest Higgs portal and their
  cosmological consequences},''
  \href{http://dx.doi.org/10.1007/JHEP10(2022)017}{{\em JHEP} {\bfseries 10}
  (2022) 017}, \href{http://arxiv.org/abs/2207.02230}{{\ttfamily
  arXiv:2207.02230 [hep-ph]}}.

\bibitem{Baldes:2023fsp}
I.~Baldes, M.~Dichtl, Y.~Gouttenoire, and F.~Sala, ``{Bubbletrons},''
  \href{http://arxiv.org/abs/2306.15555}{{\ttfamily arXiv:2306.15555
  [hep-ph]}}.

\bibitem{Friedlander:2020tnq}
A.~Friedlander, I.~Banta, J.~M. Cline, and D.~Tucker-Smith, ``{Wall speed and
  shape in singlet-assisted strong electroweak phase transitions},''
  \href{http://dx.doi.org/10.1103/PhysRevD.103.055020}{{\em Phys. Rev. D}
  {\bfseries 103} no.~5, (2021) 055020},
  \href{http://arxiv.org/abs/2009.14295}{{\ttfamily arXiv:2009.14295
  [hep-ph]}}.

\bibitem{Cai:2020djd}
R.-G. Cai and S.-J. Wang, ``{Effective picture of bubble expansion},''
  \href{http://dx.doi.org/10.1088/1475-7516/2021/03/096}{{\em JCAP} {\bfseries
  03} (2021) 096}, \href{http://arxiv.org/abs/2011.11451}{{\ttfamily
  arXiv:2011.11451 [astro-ph.CO]}}.

\bibitem{Cline:2021iff}
J.~M. Cline, A.~Friedlander, D.-M. He, K.~Kainulainen, B.~Laurent, and
  D.~Tucker-Smith, ``{Baryogenesis and gravity waves from a UV-completed
  electroweak phase transition},''
  \href{http://dx.doi.org/10.1103/PhysRevD.103.123529}{{\em Phys. Rev. D}
  {\bfseries 103} no.~12, (2021) 123529},
  \href{http://arxiv.org/abs/2102.12490}{{\ttfamily arXiv:2102.12490
  [hep-ph]}}.

\bibitem{Bea:2021zsu}
Y.~Bea, J.~Casalderrey-Solana, T.~Giannakopoulos, D.~Mateos,
  M.~Sanchez-Garitaonandia, and M.~Zilh\~ao, ``{Bubble wall velocity from
  holography},'' \href{http://dx.doi.org/10.1103/PhysRevD.104.L121903}{{\em
  Phys. Rev. D} {\bfseries 104} no.~12, (2021) L121903},
  \href{http://arxiv.org/abs/2104.05708}{{\ttfamily arXiv:2104.05708
  [hep-th]}}.

\bibitem{Bigazzi:2021ucw}
F.~Bigazzi, A.~Caddeo, T.~Canneti, and A.~L. Cotrone, ``{Bubble wall velocity
  at strong coupling},'' \href{http://dx.doi.org/10.1007/JHEP08(2021)090}{{\em
  JHEP} {\bfseries 08} (2021) 090},
  \href{http://arxiv.org/abs/2104.12817}{{\ttfamily arXiv:2104.12817
  [hep-ph]}}.

\bibitem{Lewicki:2021pgr}
M.~Lewicki, M.~Merchand, and M.~Zych, ``{Electroweak bubble wall expansion:
  gravitational waves and baryogenesis in Standard Model-like thermal
  plasma},'' \href{http://dx.doi.org/10.1007/JHEP02(2022)017}{{\em JHEP}
  {\bfseries 02} (2022) 017}, \href{http://arxiv.org/abs/2111.02393}{{\ttfamily
  arXiv:2111.02393 [astro-ph.CO]}}.

\bibitem{Dorsch:2021nje}
G.~C. Dorsch, S.~J. Huber, and T.~Konstandin, ``{A sonic boom in bubble wall
  friction},'' \href{http://dx.doi.org/10.1088/1475-7516/2022/04/010}{{\em
  JCAP} {\bfseries 04} no.~04, (2022) 010},
  \href{http://arxiv.org/abs/2112.12548}{{\ttfamily arXiv:2112.12548
  [hep-ph]}}.

\bibitem{DeCurtis:2022hlx}
S.~De~Curtis, L.~D. Rose, A.~Guiggiani, A.~G. Muyor, and G.~Panico, ``{Bubble
  wall dynamics at the electroweak phase transition},''
  \href{http://dx.doi.org/10.1007/JHEP03(2022)163}{{\em JHEP} {\bfseries 03}
  (2022) 163}, \href{http://arxiv.org/abs/2201.08220}{{\ttfamily
  arXiv:2201.08220 [hep-ph]}}.

\bibitem{Lewicki:2022nba}
M.~Lewicki, V.~Vaskonen, and H.~Veerm\"ae, ``{Bubble dynamics in fluids with
  N-body simulations},''
  \href{http://dx.doi.org/10.1103/PhysRevD.106.103501}{{\em Phys. Rev. D}
  {\bfseries 106} no.~10, (2022) 103501},
  \href{http://arxiv.org/abs/2205.05667}{{\ttfamily arXiv:2205.05667
  [astro-ph.CO]}}.

\bibitem{Ai:2022kqm}
W.-Y. Ai, J.~S. Cruz, B.~Garbrecht, and C.~Tamarit, ``{Instability of bubble
  expansion at zero temperature},''
  \href{http://dx.doi.org/10.1103/PhysRevD.107.036014}{{\em Phys. Rev. D}
  {\bfseries 107} no.~3, (2023) 036014},
  \href{http://arxiv.org/abs/2209.00639}{{\ttfamily arXiv:2209.00639
  [hep-th]}}.

\bibitem{GarciaGarcia:2022yqb}
I.~Garcia~Garcia, G.~Koszegi, and R.~Petrossian-Byrne, ``{Reflections on Bubble
  Walls},'' \href{http://arxiv.org/abs/2212.10572}{{\ttfamily arXiv:2212.10572
  [hep-ph]}}.

\bibitem{LiLi:2023dlc}
L.~Li, S.-J. Wang, and Z.-Y. Yuwen, ``{Bubble expansion at strong coupling},''
  \href{http://arxiv.org/abs/2302.10042}{{\ttfamily arXiv:2302.10042
  [hep-th]}}.

\bibitem{Krajewski:2023clt}
T.~Krajewski, M.~Lewicki, and M.~Zych, ``{Hydrodynamical constraints on bubble
  wall velocity},'' \href{http://arxiv.org/abs/2303.18216}{{\ttfamily
  arXiv:2303.18216 [astro-ph.CO]}}.

\bibitem{Giombi:2023jqq}
L.~Giombi and M.~Hindmarsh, ``{General relativistic bubble growth in
  cosmological phase transitions},''
  \href{http://arxiv.org/abs/2307.12080}{{\ttfamily arXiv:2307.12080
  [astro-ph.CO]}}.

\bibitem{Paz:1990sd}
J.~P. Paz, ``{Dissipative effects during the oscillations around a true
  vacuum},'' \href{http://dx.doi.org/10.1103/PhysRevD.42.529}{{\em Phys. Rev.
  D} {\bfseries 42} (1990) 529--542}.

\bibitem{Boyanovsky:1994me}
D.~Boyanovsky, H.~J. de~Vega, R.~Holman, D.~S. Lee, and A.~Singh,
  ``{Dissipation via particle production in scalar field theories},''
  \href{http://dx.doi.org/10.1103/PhysRevD.51.4419}{{\em Phys. Rev. D}
  {\bfseries 51} (1995) 4419--4444},
  \href{http://arxiv.org/abs/hep-ph/9408214}{{\ttfamily arXiv:hep-ph/9408214}}.

\bibitem{Yokoyama:2004pf}
J.~Yokoyama, ``{Fate of oscillating scalar fields in the thermal bath and their
  cosmological implications},''
  \href{http://dx.doi.org/10.1103/PhysRevD.70.103511}{{\em Phys. Rev. D}
  {\bfseries 70} (2004) 103511},
  \href{http://arxiv.org/abs/hep-ph/0406072}{{\ttfamily arXiv:hep-ph/0406072}}.

\bibitem{Mukaida:2012qn}
K.~Mukaida and K.~Nakayama, ``{Dynamics of oscillating scalar field in thermal
  environment},'' \href{http://dx.doi.org/10.1088/1475-7516/2013/01/017}{{\em
  JCAP} {\bfseries 01} (2013) 017},
  \href{http://arxiv.org/abs/1208.3399}{{\ttfamily arXiv:1208.3399 [hep-ph]}}.

\bibitem{Mukaida:2013xxa}
K.~Mukaida, K.~Nakayama, and M.~Takimoto, ``{Fate of $Z_2$ Symmetric Scalar
  Field},'' \href{http://dx.doi.org/10.1007/JHEP12(2013)053}{{\em JHEP}
  {\bfseries 12} (2013) 053}, \href{http://arxiv.org/abs/1308.4394}{{\ttfamily
  arXiv:1308.4394 [hep-ph]}}.

\bibitem{Ai:2021gtg}
W.-Y. Ai, M.~Drewes, D.~Glavan, and J.~Hajer, ``{Oscillating scalar dissipating
  in a medium},'' \href{http://dx.doi.org/10.1007/JHEP11(2021)160}{{\em JHEP}
  {\bfseries 11} (2021) 160}, \href{http://arxiv.org/abs/2108.00254}{{\ttfamily
  arXiv:2108.00254 [hep-ph]}}.

\bibitem{Wang:2022mvv}
Z.-L. Wang and W.-Y. Ai, ``{Dissipation of oscillating scalar backgrounds in an
  FLRW universe},'' \href{http://dx.doi.org/10.1007/JHEP11(2022)075}{{\em JHEP}
  {\bfseries 11} (2022) 075}, \href{http://arxiv.org/abs/2202.08218}{{\ttfamily
  arXiv:2202.08218 [hep-ph]}}.

\bibitem{Ai:2023ahr}
W.-Y. Ai and Z.-L. Wang, ``{Fate of homogeneous $Z_2$-symmetric scalar
  condensates},'' \href{http://arxiv.org/abs/2307.14811}{{\ttfamily
  arXiv:2307.14811 [hep-ph]}}.

\bibitem{Watkins:1991zt}
R.~Watkins and L.~M. Widrow, ``{Aspects of reheating in first order
  inflation},'' \href{http://dx.doi.org/10.1016/0550-3213(92)90362-F}{{\em
  Nucl. Phys. B} {\bfseries 374} (1992) 446--468}.

\bibitem{Konstandin:2011ds}
T.~Konstandin and G.~Servant, ``{Natural Cold Baryogenesis from Strongly
  Interacting Electroweak Symmetry Breaking},''
  \href{http://dx.doi.org/10.1088/1475-7516/2011/07/024}{{\em JCAP} {\bfseries
  07} (2011) 024}, \href{http://arxiv.org/abs/1104.4793}{{\ttfamily
  arXiv:1104.4793 [hep-ph]}}.

\bibitem{Falkowski:2012fb}
A.~Falkowski and J.~M. No, ``{Non-thermal Dark Matter Production from the
  Electroweak Phase Transition: Multi-TeV WIMPs and 'Baby-Zillas'},''
  \href{http://dx.doi.org/10.1007/JHEP02(2013)034}{{\em JHEP} {\bfseries 02}
  (2013) 034}, \href{http://arxiv.org/abs/1211.5615}{{\ttfamily arXiv:1211.5615
  [hep-ph]}}.

\bibitem{Mansour:2023fwj}
H.~Mansour and B.~Shakya, ``{On Particle Production from Phase Transition
  Bubbles},'' \href{http://arxiv.org/abs/2308.13070}{{\ttfamily
  arXiv:2308.13070 [hep-ph]}}.

\bibitem{Moreno:1998bq}
J.~M. Moreno, M.~Quiros, and M.~Seco, ``{Bubbles in the supersymmetric standard
  model},'' \href{http://dx.doi.org/10.1016/S0550-3213(98)00283-1}{{\em Nucl.
  Phys. B} {\bfseries 526} (1998) 489--500},
  \href{http://arxiv.org/abs/hep-ph/9801272}{{\ttfamily arXiv:hep-ph/9801272}}.

\bibitem{Baldes:2022oev}
I.~Baldes, Y.~Gouttenoire, and F.~Sala, ``{Hot and heavy dark matter from a
  weak scale phase transition},''
  \href{http://dx.doi.org/10.21468/SciPostPhys.14.3.033}{{\em SciPost Phys.}
  {\bfseries 14} (2023) 033}, \href{http://arxiv.org/abs/2207.05096}{{\ttfamily
  arXiv:2207.05096 [hep-ph]}}.

\end{thebibliography}\endgroup

\end{document}